\documentclass{emulateapj}
\usepackage[dvipsnames]{xcolor}
\usepackage{graphicx}
\usepackage{amsmath}
\usepackage{amssymb}
\usepackage{color}
\usepackage{verbatim}
\usepackage{enumerate}
\usepackage{multirow}
\usepackage[citecolor=teal]{hyperref}
\usepackage{bm}
\usepackage{enumitem}
\usepackage{calc}

\hypersetup{
    colorlinks=true,
    linkcolor=teal,
    filecolor=magenta,
    urlcolor=teal,
}

\newcommand{\mjy}{{\rm mJy}\ }
\newcommand{\sqdeg}{{\rm deg}^{2}\ }

\newcommand{\clpp}{$\hat C_{L}^{\phi\phi}$}
\newcommand{\clxx}{$\hat C_{L}^{\psi\psi}$}
\newcommand{\clpg}{$\hat C_{L}^{\phi G}$}
\newcommand*\xx{\color{black}}

\slugcomment{}

\shorttitle{\textsc{SPT+Planck} $2500\textsc{ deg}^{2}$ \textsc{lensing} }
\shortauthors{\textsc{SPT Collaboration\ 2016}}

\begin{document}

\title{A $2500\ {\rm \deg}^{2}$ CMB lensing map from combined South Pole Telescope and \emph{Planck} data}

\def\McGill{1}
\def\KICPChicago{2}
\def\PhysicsUChicago{3}
\def\KIPAC{4}
\def\Stanford{5}
\def\Davis{6}
\def\Penn{7}
\def\AAUChicago{8}
\def\FNAL{9}
\def\ArgonneHEP{10}
\def\EFIChicago{11}
\def\SLAC{12}
\def\Caltech{13}
\def\Berkeley{14}
\def\Colorado{15}
\def\ESO{16}
\def\Colphys{17}
\def\Cifar{18}
\def\Illast{19}
\def\Illphys{20}
\def\UChicago{21}
\def\LBNL{22}
\def\Arizona{23}
\def\Michigan{24}
\def\Munich{25}
\def\ExcellenceCluster{26}
\def\MPE{27}
\def\Dunlap{28}
\def\Minnesota{29}
\def\Melbourne{30}
\def\CaseWestern{31}
\def\ArtInstChicago{32}
\def\JPL{33}
\def\CfA{34}
\def\UToronto{35}
\def\BCCP{36}

\author{
  Y.~Omori\altaffilmark{\McGill},
  R.~Chown\altaffilmark{\McGill},
  G.~Simard\altaffilmark{\McGill},
  K.~T.~Story\altaffilmark{\KICPChicago,\PhysicsUChicago,\KIPAC,\Stanford},
  K.~Aylor\altaffilmark{\Davis},
  E.~J.~Baxter\altaffilmark{\Penn,\AAUChicago,\KICPChicago}  B.~A.~Benson\altaffilmark{\FNAL,\KICPChicago,\AAUChicago},
  L.~E.~Bleem\altaffilmark{\ArgonneHEP,\KICPChicago},
  J.~E.~Carlstrom\altaffilmark{\KICPChicago,\PhysicsUChicago,\ArgonneHEP,\AAUChicago,\EFIChicago},
  C.~L.~Chang\altaffilmark{\ArgonneHEP,\KICPChicago,\AAUChicago},
  H-M.~Cho\altaffilmark{\SLAC},
  T.~M.~Crawford\altaffilmark{\KICPChicago,\AAUChicago},
  A.~T.~Crites\altaffilmark{\KICPChicago,\AAUChicago,\Caltech},
  T.~de~Haan\altaffilmark{\McGill,\Berkeley},
  M.~A.~Dobbs\altaffilmark{\McGill,\Cifar},
  W.~B.~Everett\altaffilmark{\Colorado},
  E.~M.~George\altaffilmark{\Berkeley,\ESO},
  N.~W.~Halverson\altaffilmark{\Colorado,\Colphys},
  N.~L.~Harrington\altaffilmark{\Berkeley},
  G.~P.~Holder\altaffilmark{\McGill,\Cifar,\Illast,\Illphys},
  Z.~Hou\altaffilmark{\KICPChicago,\AAUChicago},
  W.~L.~Holzapfel\altaffilmark{\Berkeley},
  J.~D.~Hrubes\altaffilmark{\UChicago},
  L.~Knox\altaffilmark{\Davis},
  A.~T.~Lee\altaffilmark{\Berkeley,\LBNL},
  E.~M.~Leitch\altaffilmark{\KICPChicago,\AAUChicago},
  D.~Luong-Van\altaffilmark{\UChicago},
  A.~Manzotti\altaffilmark{\KICPChicago,\AAUChicago}, 
  D.~P.~Marrone\altaffilmark{\Arizona},
  J.~J.~McMahon\altaffilmark{\Michigan},
  S.~S.~Meyer\altaffilmark{\KICPChicago,\AAUChicago,\EFIChicago,\PhysicsUChicago},
  L.~M.~Mocanu\altaffilmark{\KICPChicago,\AAUChicago},
  J.~J.~Mohr\altaffilmark{\Munich,\ExcellenceCluster,\MPE},
  T.~Natoli\altaffilmark{\KICPChicago,\PhysicsUChicago,\Dunlap},
  S.~Padin\altaffilmark{\KICPChicago,\AAUChicago},
  C.~Pryke\altaffilmark{\Minnesota},
  C.~L.~Reichardt\altaffilmark{\Berkeley,\Melbourne},
  J.~E.~Ruhl\altaffilmark{\CaseWestern},
  J.~T.~Sayre\altaffilmark{\CaseWestern,\Colorado},
  K.~K.~Schaffer\altaffilmark{\KICPChicago,\EFIChicago,\ArtInstChicago},
  E.~Shirokoff\altaffilmark{\Berkeley,\KICPChicago,\AAUChicago}, 
  Z.~Staniszewski\altaffilmark{\CaseWestern,\JPL},
  A.~A.~Stark\altaffilmark{\CfA},
  K.~Vanderlinde\altaffilmark{\Dunlap,\UToronto},
  J.~D.~Vieira\altaffilmark{\Illast,\Illphys}, 
  R.~Williamson\altaffilmark{\KICPChicago,\AAUChicago}, and
  O.~Zahn\altaffilmark{\BCCP} 
  }

\altaffiltext{\McGill}{Department of Physics and McGill Space Institute, McGill University, Montreal, Quebec H3A 2T8, Canada}
\altaffiltext{\KICPChicago}{Kavli Institute for Cosmological Physics, University of Chicago, Chicago, IL, USA 60637}
\altaffiltext{\PhysicsUChicago}{Department of Physics, University of Chicago, Chicago, IL, USA 60637}
\altaffiltext{\KIPAC}{Kavli Institute for Particle Astrophysics and Cosmology, Stanford University, 452 Lomita Mall, Stanford, CA 94305}
\altaffiltext{\Stanford}{Dept. of Physics, Stanford University, 382 Via Pueblo Mall, Stanford, CA 94305}
\altaffiltext{\Davis}{Department of Physics, University of California, Davis, CA, USA 95616}
\altaffiltext{\Penn}{Center for Particle Cosmology, Department of Physics and Astronomy, University of Pennsylvania, Philadelphia, PA,  USA 19104} 
\altaffiltext{\AAUChicago}{Department of Astronomy and Astrophysics, University of Chicago, Chicago, IL, USA 60637}
\altaffiltext{\FNAL}{Fermi National Accelerator Laboratory, MS209, P.O. Box 500, Batavia, IL 60510}
\altaffiltext{\ArgonneHEP}{High Energy Physics Division, Argonne National Laboratory, Argonne, IL, USA 60439}
\altaffiltext{\EFIChicago}{Enrico Fermi Institute, University of Chicago, Chicago, IL, USA 60637}
\altaffiltext{\SLAC}{SLAC National Accelerator Laboratory, 2575 Sand Hill Road, Menlo Park, CA 94025}
\altaffiltext{\Caltech}{California Institute of Technology, Pasadena, CA, USA 91125}
\altaffiltext{\Berkeley}{Department of Physics, University of California, Berkeley, CA, USA 94720}
\altaffiltext{\Colorado}{Center for Astrophysics and Space Astronomy, Department of Astrophysical and Planetary Sciences, University of Colorado, Boulder, CO, 80309}
\altaffiltext{\ESO}{European Southern Observatory, Karl-Schwarzschild-Stra{\ss}e 2, 85748 Garching, Germany}
\altaffiltext{\Colphys}{Department of Physics, University of Colorado, Boulder, CO, 80309}
\altaffiltext{\Cifar}{Canadian Institute for Advanced Research, CIFAR Program in Cosmology and Gravity, Toronto, ON, M5G 1Z8, Canada}
\altaffiltext{\Illast}{Astronomy Department, University of Illinois at Urbana-Champaign, 1002 W. Green Street, Urbana, IL 61801, USA}
\altaffiltext{\Illphys}{Department of Physics, University of Illinois Urbana-Champaign, 1110 W. Green Street, Urbana, IL 61801, USA}
\altaffiltext{\UChicago}{University of Chicago, Chicago, IL, USA 60637}
\altaffiltext{\LBNL}{Physics Division, Lawrence Berkeley National Laboratory, Berkeley, CA, USA 94720}
\altaffiltext{\Arizona}{Steward Observatory, University of Arizona, 933 North Cherry Avenue, Tucson, AZ 85721}
\altaffiltext{\Michigan}{Department of Physics, University of Michigan, Ann  Arbor, MI, USA 48109}
\altaffiltext{\Munich}{Faculty of Physics, Ludwig-Maximilians-Universit\"{a}t, 81679 M\"{u}nchen, Germany}
\altaffiltext{\ExcellenceCluster}{Excellence Cluster Universe, 85748 Garching, Germany}
\altaffiltext{\MPE}{Max-Planck-Institut f\"{u}r extraterrestrische Physik, 85748 Garching, Germany}
\altaffiltext{\Dunlap}{Dunlap Institute for Astronomy \& Astrophysics, University of Toronto, 50 St George St, Toronto, ON, M5S 3H4, Canada}
\altaffiltext{\Minnesota}{Department of Physics, University of Minnesota, Minneapolis, MN, USA 55455}
\altaffiltext{\Melbourne}{School of Physics, University of Melbourne, Parkville, VIC 3010, Australia}
\altaffiltext{\CaseWestern}{Physics Department, Center for Education and Research in Cosmology and Astrophysics, Case Western Reserve University,Cleveland, OH, USA 44106}
\altaffiltext{\ArtInstChicago}{Liberal Arts Department, School of the Art Institute of Chicago, Chicago, IL, USA 60603}
\altaffiltext{\JPL}{Jet Propulsion Laboratory, California Institute of Technology, Pasadena, CA 91109, USA}
\altaffiltext{\CfA}{Harvard-Smithsonian Center for Astrophysics, Cambridge, MA, USA 02138}
\altaffiltext{\UToronto}{Department of Astronomy \& Astrophysics, University of Toronto, 50 St George St, Toronto, ON, M5S 3H4, Canada}
\altaffiltext{\BCCP}{Berkeley Center for Cosmological Physics, Department of Physics, University of California, and Lawrence Berkeley National Labs, Berkeley, CA, USA 94720}

\begin{abstract}

We present a cosmic microwave background (CMB) lensing map produced from a linear combination of South Pole Telescope (SPT) and \emph{Planck} temperature data.
The 150 GHz temperature data from the $2500\ {\rm deg}^{2}$ SPT-SZ survey  is combined with the \emph{Planck} 143 GHz data in harmonic space, to obtain a temperature map that has a broader $\ell$ coverage and less noise than either individual map.
Using a quadratic estimator technique on this combined temperature map, we produce a map of the   gravitational lensing potential projected along the line of sight.
We measure the auto-spectrum of the lensing potential $C_{L}^{\phi\phi}$, and compare it to the theoretical prediction for a $\Lambda$CDM cosmology consistent with the \emph{Planck} 2015 data set, finding a best-fit amplitude of $0.95_{-0.06}^{+0.06}({\rm Stat.})\! _{-0.01}^{+0.01}({\rm Sys.})$.
The null hypothesis of no lensing is rejected at a significance of $24\,\sigma$. 
One important use of such a lensing potential map is in cross-correlations with other dark matter tracers. 
We demonstrate this cross-correlation in practice by calculating the cross-spectrum, $C_{L}^{\phi G}$, between the SPT+\emph{Planck} lensing map and Wide-field Infrared Survey Explorer (\emph{WISE}) galaxies. 
We fit $C_{L}^{\phi G}$ to a power law of the form $p_{L}=a(L/L_{0})^{-b}$ with $a=2.15 \times 10^{-8}$, $b=1.35$, $L_{0}=490$, and find $\eta^{\phi G}=0.94^{+0.04}_{-0.04}$, which is marginally lower, but in good agreement with $\eta^{\phi G}=1.00^{+0.02}_{-0.01}$, the best-fit amplitude for the cross-correlation of \emph{Planck}-2015 CMB lensing and \emph{WISE} galaxies over $\sim67\%$ of the sky. 
The lensing potential map presented here will be used for cross-correlation studies with the Dark Energy Survey (DES), whose footprint nearly completely covers the SPT $2500\ {\rm deg}^2$ field.
\end{abstract}

\keywords{Gravitational lensing : general ---}

\section{Introduction}\label{sec:intro}

Mapping the distribution of matter in the Universe is one of the primary goals of modern cosmology.
In the currently favored cosmological paradigm, quantum fluctuations in the extremely early Universe
were stretched to macroscopic wavelengths during the epoch of inflation and subsequently grew
under the influence of gravity, eventually creating all the structure we observe in the Universe today.
Observations of the distribution of matter in the local and distant Universe can thus inform our
understanding of the origin of fluctuations and how they grew as a function of time. The fluctuation
amplitude at redshift $z \sim 1000$ (roughly 400,000 years after inflation) is already well
constrained by measurements of the primary CMB temperature and polarization anisotropy power spectra
(e.g \citealt{planck15-11}). Measurements of structure from $z \sim 1000$ to today primarily constrain
the efficiency with which fluctuations have grown as a function of wavelength and redshift,
i.e., the cosmic growth function \citep[e.g.][]{huterer15}. Among the most interesting and important
applications of measuring the growth function are measuring the masses of the neutrinos
by detecting their influence on growth as a function of physical wavelength \citep[e.g.][]{abazajian15b},
and distinguishing between dark energy and modified gravity as the cause of cosmic acceleration \citep[e.g.][]{weinberg13,huterer15}.

Traditionally, measurements of cosmic structure have used light (or its absence) as a proxy for matter. Galaxy and quasar surveys, measurements of the Lyman-$\alpha$ forest, and optical, X-ray, and Sunyaev-Zel'dovich effect surveys for clusters of galaxies all fundamentally rely on radiation emitted, absorbed, or distorted by collapsed objects as a tracer of the underlying matter field.
Analyses of the statistics of these tracers assume a relation between the observed property of the collapsed objects and the mass of those objects, and between the statistics of the collapsed objects and those of the underlying matter field. These assumptions are a source of systematic uncertainty in analyses of tracer populations.

A new frontier in the measurement of cosmic structure is the use of gravitational lensing to measure the matter distribution directly.
While strong gravitational lensing is useful for studying individual distant objects and small regions of sky in detail, weak lensing is a very promising avenue for measuring the large-scale distribution of matter. Weak lensing measurements of the matter distribution can be made using the coherent distortions of the shapes of galaxies measured at optical wavelengths, often referred to as cosmic shear \citep[e.g.,][]{kaiser92,bernardeau97a,kilbinger15}, and using the weak gravitational lensing of the cosmic microwave background (CMB, e.g., \citealt{lewis06}).
Lensing of the CMB has some key advantages over cosmic shear and other lensing probes, in that the CMB is a single well-localized source at a precisely known redshift, and the underlying statistics of the CMB are well
characterized and known to be very close to Gaussian.
These characteristics make reconstructing maps of the mass or gravitational potential responsible for CMB lensing a comparatively straightforward process.

CMB lensing encodes the effect of all matter fluctuations along the line of sight between the observer and the CMB last-scattering surface at $z \sim 1100$, though the CMB is most efficiently lensed by matter at $0.5 \lesssim z \lesssim 3$ \citep[e.g.,][]{zaldarriaga99}.
This wide and well-estimated redshift kernel makes CMB lensing an ideal probe to investigate phenomena such as massive neutrinos that affect the shape of the matter power spectrum.
Although there is no information about the redshift dependence of growth inherent to CMB lensing maps,
correlating CMB lensing maps with other tracers of large-scale structure that do have redshift information---optical galaxy surveys in particular---can provide interesting constraints on dark energy and modified gravity \citep[e.g.,][]{giannantonio16,kirk16}.

The ideal CMB lensing map for cross-correlation studies would thus have high signal-to-noise (S/N) on all angular modes of interest, large sky coverage, and significant overlap with galaxy surveys.
The highest total S/N of any measurement of CMB lensing comes from the
lensing map produced by the \emph{Planck} collaboration, which covers nearly the full sky but with low S/N per mode, and only at scales below tens of degrees.
Data from ground-based telescopes such as the South Pole Telescope (SPT), the Atacama Cosmology Telescope (ACT), and POLARBEAR have been used to produce lensing maps with improved S/N on smaller scales over much smaller sky fractions \citep{das11,vanengelen12,polarbear13b,story15,sherwin16}.

In this work, we present a CMB lensing map over 2500 square degrees---the largest lensing map yet produced using a high-resolution ground-based experiment---and with nearly 100\% overlap with the Dark Energy Survey
(DES)\footnote{http://www.darkenergysurvey.org/} optical galaxy survey. This map is derived from the optimal combination of SPT data from the 2500 $\mathrm{deg}^2$ SPT-SZ survey \citep{story13} with \emph{Planck} temperature data covering the same patch of sky.
We expect this map to be particularly useful for cross-correlation analyses, particularly with DES, and the optimal combination of SPT and \emph{Planck} data should allow nearly maximal exploitation of these three powerful data sets. 
The map will be made publicly available in an upcoming SPT data release.

As validations and consistency checks, 
we calculate the power spectrum of this map and its cross-correlation with infrared-selected galaxies
from the Wide-field Infrared Survey Explorer (\emph{WISE}) survey.
This paper has an accompanying paper (Simard et al., in preparation), which analyzes the best-fit cosmological parameters and explores the physical implications, using the CMB lensing map presented in this paper.

This paper is laid out as follows: we first provide an overview of the method for reconstruction of the CMB weak gravitational lensing potential in Section \ref{sec:theory}; SPT and $Planck$ 
 data are described in Section \ref{sec:data}; sky simulations used for generating mock reconstructed lensing potentials are described in Section \ref{sec:sims}; the process of combining SPT and $Planck$ temperature maps is outlined in Section \ref{sec:methods}; the resulting gravitational lensing map and angular power spectra are presented in Section \ref{sec:results}; systematic checks and potential sources of errors are described in Section \ref{sec:sys}; implications and discussions of the results are given in Section \ref{sec:discussions}.

{\xx Throughout this paper, we assume
a spatially flat $\Lambda$CDM \emph{Planck} 2015 cosmology}\footnote{base\_plikHM\_TT\_lowTEB\_lensing} \citep{planck15-15} with parameters $\Omega_{\rm b}h^{2}=0.022$, $\Omega_{\rm c}h^{2}=0.12$, $\Omega_{\rm m}=0.31$, $H_{0}=100\ h {\rm km\ s}^{-1} {\rm Mpc}^{-1}$ with $h = 0.68$, power spectrum of primordial curvature perturbations with an amplitude (at $k = 0.05\ {\rm Mpc}^{-1}$) $A_{\rm s} = 2.1\times10^{-9}$ and spectral index $n_{\rm s} = 0.97$, amplitude of the (linear) power spectrum on the scale of $8\ h^{-1}{\rm Mpc}\ \sigma_{8}=0.82$,  optical depth to reionization $\tau = 0.067$, and we assume one massive neutrino with a 0.06 eV mass. We use the subscript ``fid" to denote a quantity calculated from the best-fit \emph{Planck} cosmology.

In discussing angular multipole moments, we use $\ell,m$ to denote CMB temperature and $L,M$ for multipole moments of the lensing field. {\xx Furthermore, we will denote the filtered lensing potentials as $\bar \phi$, the estimated lensing potential as $\hat \phi$, the \emph{masked} estimated lensing potential as $\tilde\phi$, and the true lensing potential as ordinary $\phi$.}

\section{Theory}\label{sec:theory}
As photons traverse the Universe from the last scattering surface, their paths are altered by the gravitational fields induced by large-scale structure, in a process known as gravitational lensing. The observed lensed temperature field $T^{\rm len}(\bf \hat n)$ can be expressed in terms of the original unlensed temperature field $T^{\rm unl}({\bf \hat n})$ \citep{lewis06} as:
\begin{equation}
T^{\rm len}({\bf \hat n})=T^{\rm unl}({\bf \hat n}+\boldsymbol{\vec{\alpha}}),
\end{equation}
where ${\bf \hat n}$ is the directional vector and $\vec{\boldsymbol{\alpha}}$ is the deflection field.
For small scalar perturbations, assuming the Born approximation, the deflection field can be described as $\nabla \phi$, where $\phi$ is the projected gravitational potential:
\begin{equation}
\phi({\bf \hat n} )=-\frac{2}{c^{2}} \int_{0}^{\chi_{\rm CMB}} d\chi \frac{f_{K}(\chi_{\rm CMB}-\chi)}{f_{K}(\chi_{\rm CMB})f_{K}(\chi)}\Psi(\chi{\bf \hat n;\eta_{0}-\chi}),
\end{equation}
where $\chi$ is the comoving distance, $f_{K}$ is the comoving angular-diameter distance and $\Psi$ is the 3D gravitational potential at conformal
distance $\chi$ along the direction ${\bf \hat n}$, and $\eta_{0}-\chi$ is the conformal time \citep{lewis06}. The deflection field is defined as $\nabla \phi({\bf \hat n})$, and the divergence of this field gives the convergence $\kappa$:
\begin{equation}
\kappa=-\frac{1}{2}\nabla^{2}\phi.
\end{equation}
In general, the deflection component additionally could contain a curl term $\nabla\times\psi$; to first order, this term is not expected \citep{namikawa12}. Although there exist mechanisms such as gravitational waves \citep{cooray05} that will generate such a mode, the amplitude is expected to be comparatively small and can be neglected. Therefore, the presence of curl modes most likely
signifies systematic errors introduced in the lensing reconstruction process, and curl modes can be used as a null test.

For a fixed lensing potential $\phi$ and multiple realizations of the unlensed CMB, the observed temperature fluctuations will have a Gaussian distribution.
Lensing introduces correlations between previously uncorrelated modes, which introduces off-diagonal terms into the harmonic-space covariance of the CMB. Expanding the temperature field in spherical harmonics, at first order in the lensing potential, the total contribution to the off-diagonal terms is  \citep{okamoto03}:
\begin{align}
\Delta\langle T_{\ell_{1},m_{1}}&T_{\ell_{2},m_{2}}\rangle\nonumber\\
&=\sum_{LM}(-1)^{M}
 \begin{pmatrix}
  \ell_{1} & \ell_{2} & L \\
  m_{1} & m_{2} & -M
 \end{pmatrix}
 W_{\ell_{1}\ell_{2}L}^{\phi}\phi_{LM},
\end{align}
where $T_{\ell m}$ are the spherical harmonic expansion coefficients of the temperature field, the term in the bracket is the Wigner $3$-$j$ symbol, and the weight function $W_{\ell_{1}\ell_{2}L}^{\phi}$ is given by:
\begin{align}
&W_{\ell_{1}\ell_{2}L}^{\phi}=-\mathstrut{\sqrt\frac{(2\ell_{1}+1)(2\ell_{2}+1)(2L+1)}{4\pi}}\nonumber\\
&\times C_{\ell_{1}}^{TT}\left(\frac{1+(-1)^{\ell_{1}+\ell_{2}+L}}{2}\right) \begin{pmatrix}
  \ell_{1} & \ell_{2} & L \\
  1 & 0 & -1\nonumber \\
 \end{pmatrix}\\
&\times \sqrt{L(L+1)\ell_{1}(\ell_{1}+1)}+(\ell_{1}\leftrightarrow\ell_{2}),
\end{align}
{\xx where $C_{\ell}^{TT}$ is the power spectrum of the lensed CMB and $\leftrightarrow$ denotes an additional term that is a replication of the first term, but with $\ell_{1}$ and $\ell_{2} $ switched.}
Estimating this covariance makes it possible to reconstruct the lensing potential. A  formally optimal estimator at first order can be written as \citep{okamoto03}:
\begin{align}
\bar{\phi}_{LM}=\frac{(-1)^{M}}{2}\sum_{\ell_{1} m_{1} \ell_{2} m_{2}}
&\begin{pmatrix}
  \ell_{1} & \ell_{2} & L \\
  m_{1} & m_{2} & -M \\
 \end{pmatrix}\nonumber \\
&W_{\ell_{1}\ell_{2}L}^{\phi}\bar{T}_{\ell_{1}m_{1}}\bar{T}_{\ell_{2} m_{2}}  \ .
\end{align}
The overbar on $T_{\ell m}$ denotes that the temperature fields have been filtered, which we discuss in Section \ref{sec:methods_filtering}. To account for this filtering, $\bar\phi_{LM}$ is normalized using a response function $\mathcal{R}_{LM}^{\bar\phi\bar\phi}$:
\begin{equation}\label{eq:response}
\hat\phi_{LM}=\frac{1}{\mathcal R_{LM}^{\bar\phi\bar\phi}}(\bar\phi_{LM}-\bar\phi_{LM}^{{\rm MF}}),
\end{equation}
where $\bar\phi^{{\rm MF}}$ is the ``mean-field" bias, which originates from any analysis steps that introduce statistical anisotropy (such as inhomogeneous noise and mode coupling induced when a mask of complex geometry is applied).
We calculate the mean-field as the average reconstructed $\bar \phi$ from 198 simulations.

We utilize the full-sky routines implemented in the {\tt quicklens}\footnote{\url{https://github.com/dhanson/quicklens}} package to compute the quadratic building blocks. The package takes advantage of the separability of the weight functions which leads to estimators that can be evaluated efficiently \citep{planck15-15}.

\begin{figure*}
\begin{center}
\includegraphics[width=0.93\textwidth]{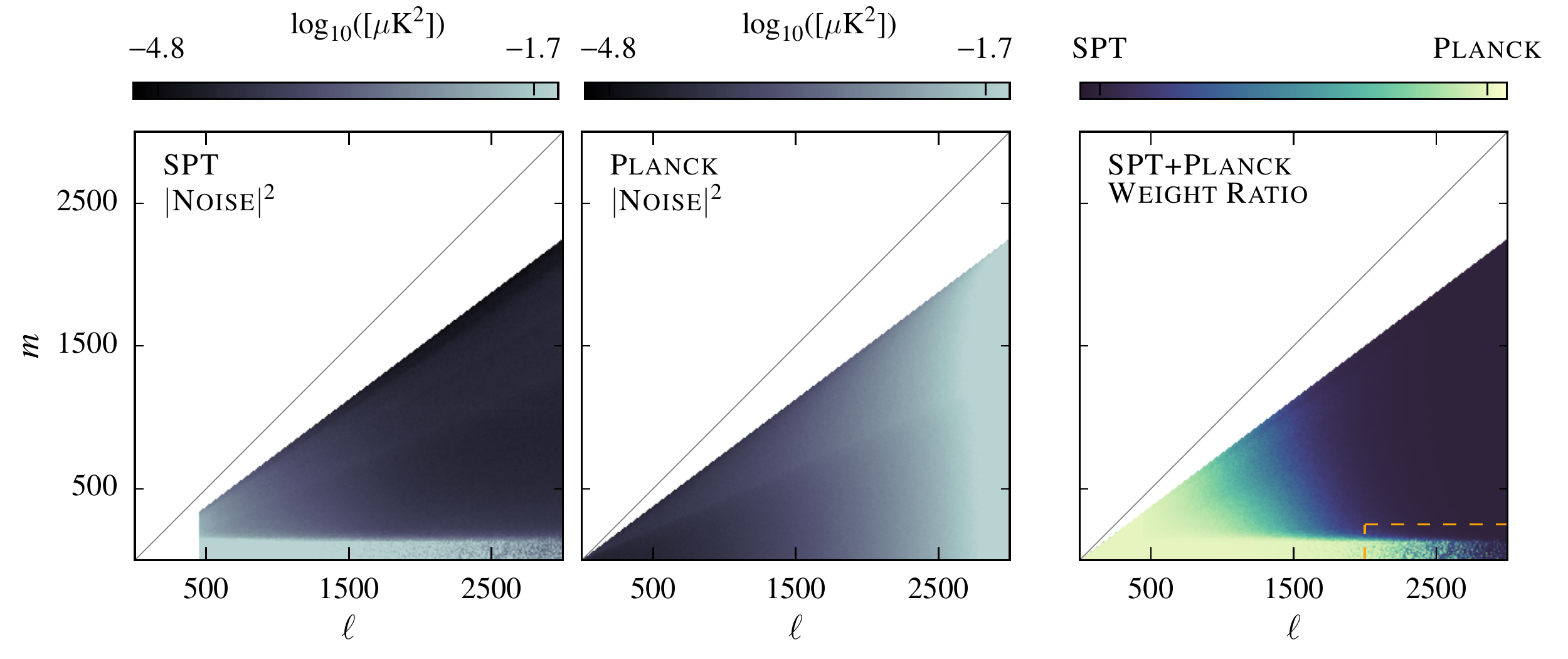}
\caption{Noise characteristics of SPT and \emph{Planck} data and the ratio of weights used in combining the data, all shown on $(\ell,m)$ grids. {\bf Left:} Transfer-function-deconvolved SPT noise. The noisy low $m$ stripe is due to the scanning strategy of SPT. {\bf Center:} Beam-deconvolved \emph{Planck} noise. {\bf Right:} Ratio of weights for SPT and Planck on a linear scale ranging from light yellow (all \emph{Planck}) to dark blue (all SPT).
In all the panels, high $m$ modes ($m>0.75 \ell$) where the values are small due to the mask, have been zeroed out to retain the scale.}
\label{fig:weights}
\end{center}
\end{figure*}

\section{Data}\label{sec:data}
\subsection{SPT Data}\label{sec:data_spt}
The South Pole Telescope (SPT) \citep{carlstrom11} is a 10-meter telescope
located at the National Science Foundation Amundsen-Scott South
Pole Station in Antarctica. From 2008-2011, the telescope was used to conduct the SPT-SZ survey, a survey
spanning a contiguous area of approximately 2500~$\sqdeg$ \citep{story13}.
The survey footprint extends from $20^{h}$ to $7^{h}$ in right ascension (R.A.)
and from $-65^\circ$ to $-40^\circ$ in declination.
The total area is divided into 19 different fields, with roughly
1 degree of overlapping coverage at the field boundaries.
These fields were observed at three frequency bands centered
at roughly 95, 150, and 220~GHz. For this analysis, we will exclusively work with the 150\ GHz data.
The 150~GHz beam has a shape similar to a Gaussian with a full-width at half maximum (FWHM) of $1.2'$. The typical noise level of SPT-SZ maps at 150~GHz is $18\ \mu K$-${\rm arcmin}$, with small variations across the different fields.
For each field, maps of individual observations were made by combining a large number of scans performed along azimuth, with small steps in elevation
in between scans.
These observations are then co-added into a single map of each field.

Due to the geographic location of the telescope, scans in azimuth correspond to scans along lines of fixed declination in equatorial coordinates.
Any low-frequency noise uncorrelated between detectors results in elevated noise levels along the scan direction, which for this observing strategy maps directly into noise at low $m$, independent of $\ell$.
Meanwhile, correlated low-frequency noise (such as from long-wavelength atmospheric fluctuations) results in isotropic large-angular-scale noise, or noise at low $\ell$, independent of $m$.
SPT data are filtered in the time domain before being projected onto maps, significantly suppressing these low-$m$ and low-$\ell$ modes (see \citealt{schaffer11} for details).
These low-$m$ and low-$\ell$ modes are poorly characterized in SPT data alone, motivating us to incorporate information from \emph{Planck} (see Figure \ref{fig:weights}). The SPT data map and the noise maps are calibrated to match the \emph{Planck} 143 GHz data using the cross-spectrum between the two data sets \citep{hou17}. We evaluate the sensitivity of the lensing map to the exact value of this calibration factor in Section \ref{sec:sys_spt_calb}.

Maps for each of the fields are produced at a native resolution of 1 arcminute in the zenithal equal-area projection and are projected onto a single {\tt HEALPix}\footnote{\url{http://healpix.sourceforge.net
}}\citep{gorski05} map of $N_\mathrm{side}=8192$. Pixels near the field boundaries that are included in multiple fields are combined using inverse variance weights. At this resolution, if present, pixelization and projection artifacts will affect $\ell>10000$, which is far beyond the $\ell_{\rm max}=3000$ that we consider in the baseline lensing analysis here.

\subsection{Planck Data}
{\xx The \emph{Planck} satellite, launched in 2009 by the European Space Agency \citep{planck15-1}, was used  to observe the millimeter sky in nine frequency bands ranging from 30 to 857~GHz.
It achieved better resolution, higher sensitivity, and a wider range of frequencies than its predecessor, the Wilkinson Microwave Anisotropy Probe (\emph{WMAP}, \citealt{bennett03a}).
In this work, we use the publicly available \emph{Planck} 143~GHz map\footnote{\url{http://irsa.ipac.caltech.edu/data/Planck/release\_2/all-sky-maps/maps/HFI\_SkyMap\_143\_2048\_R2.02\_full.fits}} and beam\footnote{\url{http://irsa.ipac.caltech.edu/data/Planck/release\_2/ancillary-data/HFI\_RIMO\_Beams-100pc\_R2.00.fits}} provided in the 2015 data release \citep{planck15-8}, as the 143\,GHz frequency band is closest to the SPT-SZ 150\,GHz band. 
The 143~GHz beam can be approximated by an azimuthally symmetric Gaussian beam with a FWHM of $\sim7$ arcmin, and the instrument noise is approximately white with an RMS of $\sim30\ \mu\mathrm{K}$-arcmin \citep{planck15-8}.
}

\section{Simulations}\label{sec:sims}
Simulations of the temperature and noise maps are used to obtain key building blocks of this analysis including the SPT transfer function and the average noise $\langle |n_{\ell m}|^{2}\rangle$, which are used to define the weights used in the combining process (described in Section \ref{sec:methods}).

Simulated temperature maps consist of 3 components:
\begin{enumerate}[label=(\roman*)]
\item Lensed CMB,
\item Gaussian foregrounds: thermal Sunyaev-Zel'dovich effect (tSZ), kinetic Sunyaev-Zel'dovich effect (kSZ), cosmic infrared background (CIB), and unresolved background of faint radio sources ($F_{150}<6.4$\ mJy),
\item Individually detected point sources: radio and dusty star forming galaxies.
\end{enumerate}
For (i), lensed CMB maps are produced by running {\tt LensPix} \citep{lewis05} with $C^{TT}_{\ell,{\rm unl}}$ calculated using {\tt CAMB} \citep{lewis00} with cosmological parameters defined in Section \ref{sec:intro} as input.  
We produce maps in {\tt HEALPix} format with $N_{\rm side}=8192$ and apply a cut-off in the input spectrum at $\ell_{\rm max}=9500$.
The resulting lensed maps are consistent with the theoretically calculated lensed spectrum $C^{TT}_{\ell,{\rm len}}$ up to $\ell\sim7000$. 
For each realization of the lensing potential, we lens two background CMB maps, which yields two sets of lensed CMB maps. 
The purpose of this second set will be explained in Section \ref{sec:method_clpp}. For (ii), we add simulated Gaussian foreground components. The shapes of the tSZ and kSZ spectrum are taken from \citet{shaw10} and \citet{shaw12} models respectively, with the amplitudes calibrated to match with \citet{george15}. 
A similar procedure is followed for the CIB component using templates from \citet{george15}.
For the clustered CIB component, the spectrum is set to follow $D_\ell \propto \ell^{0.8}$ with an amplitude $D_{3000}^c = 3.46\ \mathrm{\mu K}^2$. The shot-noise or ``Poisson'' CIB power from galaxies dimmer than $6.4 \mathrm{mJy}$ is taken to be $D_{3000}^{P} = 9.16 \ {\mu {\rm K}}^2$.
For the unresolved faint radio sources, we generate random realizations using $\mathrm{d}N/\mathrm{d}S$ taken from \citet{dezotti05}, and calibrate the amplitude using SPT 150\ GHz observations.

We place point sources at the observed locations with their measured fluxes for point sources in the flux range $6.4\ \mjy < F_{150} < 50\  {\rm mJy}$ listed in the SPT point source catalog (Everett et al., in preparation).
Similarly, we add clusters with significance $\xi>4.5$ listed in \cite{bleem15b} and model the profile using a projected isothermal $\beta$ model \citep{cavaliere76} with $\beta=1$. From these inputs, we produce the simulated SPT and \emph{Planck} maps separately.

For the SPT simulations, the input {\tt HEALPix} maps are passed through a mock observing pipeline, which creates mock time-ordered data from these maps for each SPT detector, filters those data in the same manner as the real data, and creates maps using the inverse-noise weights from the real data.
The observation runs for each of the fields are co-added and the beams are deconvolved using the beam models associated with those fields. All the fields are then re-convolved with a FWHM$=1.75'$ Gaussian beam, projected onto a single {\tt HEALPix} map of $N_{\rm side}=8192$, and then stitched to produce a conjoint $2500\ \sqdeg$ map.

From the noise-free mock maps, we calculate the filter transfer function
\begin{equation}\label{eq:spt_tf}
\mathcal{Y}_{\ell m}=\frac{\langle T_{\ell m}^{\rm out}T_{\ell m}^{\rm in,*} \rangle}{\langle T_{\ell m}^{\rm in}T_{\ell m}^{\rm in,*} \rangle},
\end{equation}
 where the $T_{\ell m}$ are computed from the \emph{boundary masked} (defined in Section \ref{sec:methods}) temperature maps, and the superscripts ``out" and ``in" refer to the outputs and inputs of the mock observing pipeline. Noise maps are produced separately by taking the differences of various SPT observations, which effectively removes the signal, and leaves noise behind. We add noise maps obtained in this way to the noise-free outputs to produce realistic data-like maps.

{\xx To produce simulated \emph{Planck} maps, we simply convolve the input signal maps by
the \emph{Planck} 143 GHz beam and add noise from the 8th Full Focal Plane simulation set (FFP8) \citep{planck15-12}.
Since we expect the SPT 150\ GHz and \emph{Planck} 143\ GHz to have a similar response to foreground signal, we do not introduce additional free parameters correcting for this small difference.
}
\section{Methods}\label{sec:methods}
This section is divided into three parts: the combining of SPT and \emph{Planck} temperature maps, the $\hat \phi$ map reconstruction procedures, and the auto-spectrum calculation. Steps taken in the two sections are presented as sequential \emph{sub}-sections to illustrate the work flow.

\subsection{Combining the SPT and Planck maps}\label{sec:combining}
We form a nearly optimal combination of SPT and \emph{Planck} data by taking the
inverse-variance-weighted sum of the two data sets in harmonic space, after deconvolving
the beam and any filtering from each data set. The procedure used in this work is similar
to that in \citet{crawford16}, with certain key differences, including the use of curved-sky maps and
spherical harmonic transforms instead of flat-sky projections and fast Fourier transforms.
To avoid real-space artifacts, we apodize
the data and mask bright sources and galaxy clusters before transforming to harmonic
space.
We also mask certain noisy $\ell,m$ modes before transforming the combination
back to real space. 
Each of these steps is described in more detail below.

\subsubsection{Boundary Mask}\label{sec:combining_boundarymask}
{\xx A binary mask (with values=1 for unmasked and 0 for masked pixels) defined by the nominal SPT region ($20^{h} < {\rm R.A.} < 7^{h}$ and  $-65^{\circ} < {\rm decl.} < -40^{\circ}$) is first produced. 
The {\tt process\_mask} routine in the {\tt HEALPix} package is then used to calculate the distance from the nearest masked pixel, and this distance map is smoothed using a Gaussian beam of FWHM$=15'$. This smoothing is applied to soften the corners of the mask. 
The distance map is then used to apodize the binary mask with a Gaussian beam of FWHM$=30'$. 
This results in a mask with an effective area of $\sim 2350\ \sqdeg$ which we apply to both SPT and \emph{Planck} maps.}

\subsubsection{Masking Bright Point Source and Clusters}\label{sec:combining_psmask}
 The brightest point sources are masked prior to the combining procedure to avoid artifacts that result from applying spherical harmonic transforms on band-limited maps. 
Apertures of radius $R=6'$ and $R=9'$ are placed at the locations of point sources with $50<F_{150}<500\ {\rm mJy}$, and $F_{150}>500\ {\rm mJy}$ respectively. 
In addition, clusters above $\xi>6$ (where $\xi$ is the detection significance, as defined in \citealt{bleem15b}) are masked with an aperture of $R=6'$. 
This cluster masking threshold balances minimizing both the tSZ contamination and the masked sky area. 

\subsubsection{Forming a Combined Map}\label{sec:combining_combining}

We combine the SPT and \emph{Planck}  maps to yield a map with lower noise   at all scales.
 This is achieved by constructing a simple linear combination of the SPT and \emph{Planck} maps in spherical harmonic space, weighted by their relative noise variance for each mode $(\ell,m)$:

\begin{align}
w_{\ell m}^{\rm SPT}&=\frac{1}{\langle|n_{\ell m }^{\rm SPT}|^{2}\rangle}\label{eq:w_s}\\
w_{\ell m}^{Planck}&=\frac{1}{\langle |n_{\ell m }^{Planck}|^{2}\rangle}\label{eq:w_p}
\end{align}
\begin{equation}
T_{\ell m}^{X}=\frac{w_{\ell m}^{\rm SPT}}{w_{\ell m}^{\rm SPT}+w_{\ell m}^{Planck}} T_{\ell m}^{\rm SPT}+ \frac{w_{\ell m}^{Planck}}{w_{\ell m}^{\rm SPT}+w_{\ell m}^{Planck}} T_{\ell m}^{Planck},
\end{equation}
where $n_{\ell m}^{{\rm SPT}}$, $n_{\ell m}^{Planck}$  are the noise estimates and $w_{\ell m}^{{\rm SPT}}$, $w_{\ell m}^{Planck}$ are the normalized weights, and  $T_{\ell m}^{X}$ are the combined SPT + \emph{Planck} spherical harmonic coefficients.

 Prior to combining, we deconvolve the 2D filter transfer function obtained using equation \ref{eq:spt_tf}, and the effective beam from SPT data and estimated noise.  We also deconvolve the \emph{Planck} beam from the \emph{Planck} data and noise estimates prior to combining.

After combining, a Gaussian beam of FWHM=$1.75'$ is applied to mitigate ringing features in the resulting map for the same reasons as mentioned in Section \ref{sec:combining_psmask}.

\subsubsection{Masking Poorly Constrained Modes}\label{sec:combining_modemasking}
{\xx 
We choose to mask the modes that are poorly constrained by both SPT and \emph{Planck}. 
Due to the SPT scanning strategy, SPT's $1/f$ noise shows up in low-$m$ modes. 
At low-$\ell$, this simply means that more weight is given to the comparatively low noise \emph{Planck} measurements of these low-$\ell$, low-$m$ modes
However, neither experiment measures high-$\ell$, low-$m$ modes well. 
The features these ill-constrained modes produce when transformed back from spherical harmonic space to real space are difficult to treat in map space, so we mask especially noisy modes. 
  In our baseline analysis, we set all modes $\ell>2000$ and $m<250$ to zero (shown as the area enclosed the orange dashed lines in Figure \ref{fig:weights}). 
The effect of varying these cuts is discussed in Section  \ref{sec:sys_lmcut}}.

\subsubsection{Faint Point Source Inpainting}\label{sec:inpainting}
In addition to the bright sources and galaxy clusters masked in Section~\ref{sec:combining_psmask}, the
combined SPT-\emph{Planck} map also contains point sources with fluxes between $6.4<F_{150}<50$ mJy. These point sources are painted over in the combined map using the constrained Gaussian inpainting method \citep{hoffman91,benoitlevy13}. The correlation function used to reconstruct the masked region
is estimated from surrounding regions:
\begin{equation}
T^{\rm obs}_{i}=T^{\rm sim}_{i}+\Xi_{ ij}\Xi_{jj}^{-1}(T^{\rm obs}_{j}-T^{\rm sim}_{j}),
\end{equation}
where the matrices $\Xi_{ij}$ and $\Xi_{jj}$ represent the cross-correlation between the region inside (denoted by subscript $i$) and outside the masked region (denoted by subscript $j$), and the auto-correlation of the outer region, respectively. The elements of these matrices are estimated using the correlation function calculated from a fiducial lensed CMB spectrum, using:
\begin{equation}
w^{TT}(\theta)=\frac{2\ell+1}{4\pi}\sum_{\ell}C_{\ell,{\rm fid}}^{TT}P_{\ell}({\rm cos}(\theta)).
\end{equation}
{\xx We have evaluated the validity of this method by applying this procedure on a simulated map without point sources, and obtained a difference of $\ll$1\% in power relative to the the map without inpainting. Finally, the Gaussian beam convolved in Section \ref{sec:combining_combining} is deconvolved from the maps.}

\subsection{Reconstructing Lensing $\phi$ from the Combined Map}\label{sec:phirec}
We reconstruct the CMB lensing potential $\phi$ from the combined map using the methods laid out in Section~\ref{sec:theory}.
Each step is detailed below, including the extra complexity required to treat the anisotropic noise in the SPT data properly.

\subsubsection{Filtering}\label{sec:methods_filtering}
The combined and inpainted $T_{\ell m}$ are filtered to maximize the extracted lensing signal. We choose the filter to be the sum of the \emph{lensed} CMB spectrum, foreground components and noise:
\begin{equation}\label{eq:filter}
F_{\ell m}=\frac{1}{|T_{\ell m,{\rm fid}}|^{2}+\langle|T_{\ell m, {\rm foregrounds}}|^{2}\rangle+\langle|T_{\ell m, {\rm noise}}|^{2}\rangle},
\end{equation}
where $|T_{\ell m,{\rm fid}}|^{2}$ is an expansion of the fiducial CMB input spectrum, $\langle|T_{\ell m, {\rm foregrounds}}|^{2}\rangle$ is the average foreground power measured from simulations and $\langle|T_{\ell m, {\rm noise}}|^{2}\rangle$ is the average noise power. The purpose of this filtering process is to down-weight the contribution from noisy modes.

\subsubsection{Mean-Field and Response Function}\label{sec:methods_mf_and_response}
We produce the mean-field for a specific realization by first splitting the set of simulations into two halves, and omitting the realization that we are trying to calculate the mean-field for:
\begin{align}
\bar\phi^{\rm MF,(1)}_{LM,i}&=\langle \bar\phi_{LM,i \neq j} \rangle _{j=0\rightarrow N_{\rm sim}/2},\\
\bar\phi^{\rm MF,(2)}_{LM,i}&=\langle \bar\phi_{LM,i \neq j} \rangle _{j=N_{\rm sim}/2\rightarrow N_{\rm sim} }.
\end{align}
The motivation for splitting the set of simulations is to ensure that the auto power of the mean-field is omitted in the lensing auto-spectrum calculation.

\begin{figure}
\begin{center}
\vspace{0.5cm}\includegraphics[width=0.45\textwidth]{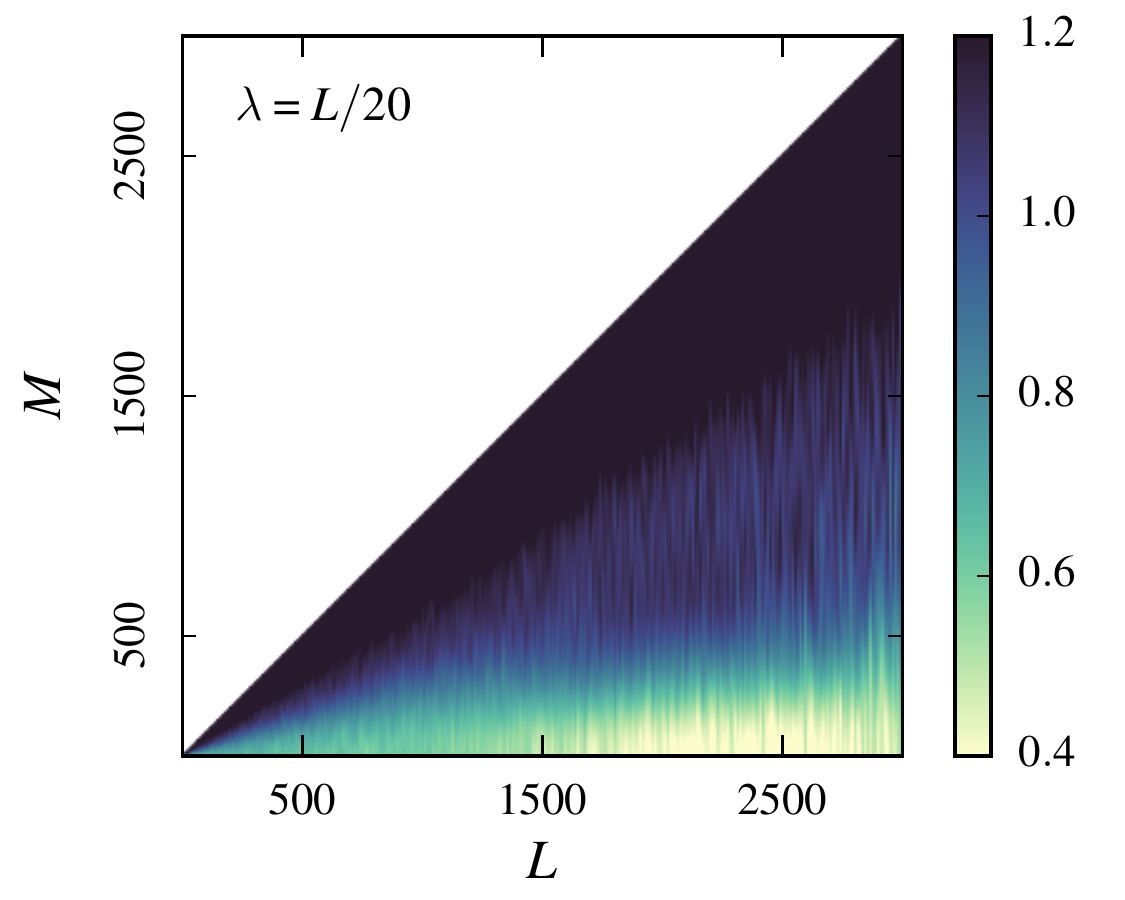}
\caption{{The ratio of smoothed 2D response function with $\lambda=L/20$ against $M$-averaged response function $\mathcal{R}_{LM}^{\phi\phi}/\mathcal{R}_{L}^{\phi\phi}$ presented on a ($L,M$) grid.} }
\label{fig:resp_2D}
\end{center}
\end{figure}

Normally the response function (Equation~\ref{eq:response}) is
assumed to be azimuthally symmetric and is only calculated as a function of $L$.
In the presence of strong $m$-dependence in the noise, as is the case for SPT, it is necessary to obtain the response function as a function of $L$ and $M$ (i.e.  $\mathcal{R}_{LM}^{\phi\phi}$). We compute this using simulations, taking the ratio of the average cross-spectrum of output (${\bar\phi}_{LM}$) and input ($\phi_{LM}$) lensing potentials to
the average auto-spectrum of the input potentials:
\begin{equation}\label{eq:resp_2D}
\mathcal{R}_{LM}^{\bar\phi\bar\phi}=\frac{\langle\phi_{LM}\bar\phi^{*}_{LM}\rangle}{\langle\phi_{LM}\phi^{*}_{LM}\rangle}.
\end{equation}
However, the response function obtained this way is rather noisy. Therefore, we apply a scale dependent Gaussian smoothing in $M$, with smoothing scale  $\lambda=L/20$. The ratio of the smoothed and an $M$-independent response function on a $(L,M)$ grid is shown in Figure \ref{fig:resp_2D}. Similar to the mean-field calculation, we split the response function into two halves, where the average is taken over the first half simulation set for (1), and over the second half for (2) in equation \ref{eq:resp_2D}. Finally, we normalize $\bar \phi$ using equation \ref{eq:response}, convert to convergence, produce maps, and apply the final analysis mask:

\begin{equation}
{\widetilde \phi}_{LM}=K^{-1}\int Y_{LM}^{*}({\bf \hat n})M({\bf \hat n})\left[\sum_{L'M'}Y_{L'M'}({\bf \hat n}) K\hat\phi_{L'M'}\right]
\end{equation}
where $K=\frac{1}{2}L(L+1)$, and $M({\bf \hat n})$ is the final analysis mask. For our baseline analysis, the final mask removes circular patches of $R=2'$ at the locations of point sources with flux $6.4<F_{150}<50\ {\rm mJy}$,  and $R=5'$ at the locations of clusters between $4.5<\xi<6$ in addition to the mask defined in Section \ref{sec:combining_psmask}.

\subsection{Calculation of the Auto-spectrum $C_{L}^{\phi\phi}$}\label{sec:method_clpp}

In this section, we will employ the notation:
\begin{equation}
C_{L}^{\widetilde\phi\widetilde\phi}[\widetilde\phi(\bar T\bar T)\widetilde\phi(\bar T\bar T)]\label{eq:cl}
\end{equation}
to explicitly show the particular $\widetilde \phi$ used to calculate the spectrum.

\begin{figure}
\begin{center}
\includegraphics[width=0.48\textwidth]{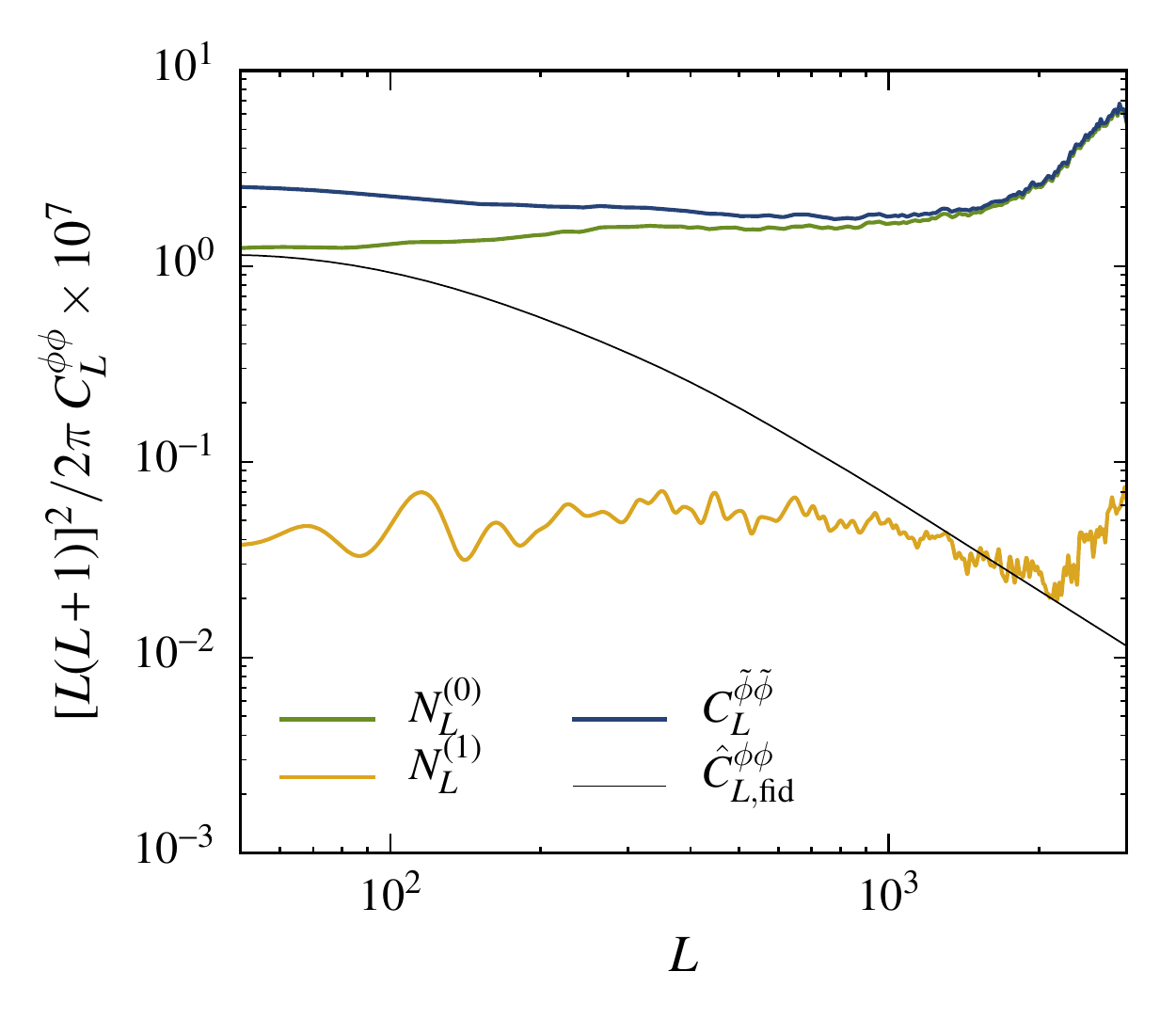}
\caption{{The amplitudes of $C_{L}^{\hat\phi\hat\phi}$ (blue), $N_{L}^{(0)}$ and $N_{L}^{(1)}$ biases (green and gold) relative to the fiducial input $\hat {C}_{L,{\rm fid}}^{\phi\phi}$} (black).}
\label{fig:biases}
\end{center}
\end{figure}

$C_{L}^{\widetilde\phi\widetilde\phi}$ calculated directly using equation \ref{eq:cl} is not equivalent to the true lensing spectrum $\hat{C}_{L}^{\phi\phi}$ since it will contain bias terms arising from correlations between the CMB and the lensing potentials. These are known as the $N^{(0)}$ bias and $N^{(1)}$ biases \citep{hu02a,kesden03,hanson11}: 
\begin{equation}
C_{L}^{\widetilde\phi\widetilde\phi}=\hat{C}_{L}^{\phi\phi}+N_{L}^{(0)}+N_{L}^{(1)}+\cdot\cdot\cdot,
\end{equation}
and the relative amplitudes are shown in Figure \ref{fig:biases}. Higher-order terms in the equation above can be neglected under the assumption that the CMB and the lensing potential follow Gaussian statistics \citep{kesden02}.

 The $N^{(0)}$ bias arises from chance correlations in the Gaussian CMB, foreground and noise, and can be written as:

\begin{align}
N_{L}^{(0)}=\biggl\langle &C_{L}^{\widetilde\phi\widetilde\phi}[\widetilde\phi^{(1)}(\bar S_{i}^{\phi_{1}}\bar S_{j}^{\phi_{2}}) \widetilde\phi^{(2)}(\bar S_{i}^{\phi_{1}}\bar S_{j}^{\phi_{2}})] +\nonumber \\
  &C_{L}^{\widetilde\phi\widetilde\phi}[\widetilde\phi^{(1)}(\bar S_{i}^{\phi_{1}}\bar S_{j}^{\phi_{2}}) \widetilde\phi^{(2)}(\bar S_{j}^{\phi_{2}}\bar S_{i}^{\phi_{1}})]
\biggl\rangle_{i,j},
\end{align}
where $\bar{S}_{i}^{\phi_{a}}$ is the $i$-th simulated temperature $T_{\ell m}$ lensed with the potential $\phi_{a}$ and filtered using equation \ref{eq:filter}. $i,j$ imply different simulation realizations. Cross-correlation of two $\widetilde\phi_{LM}$ calculated using different mean-fields and response functions (denoted by the superscripts (1), (2)) are used to ensure that the auto-spectra of these components do not affect the resulting spectrum. {\xx The $N^{(0)}$ bias is calculated from 198 mock observed simulations that contain all the foreground components (point sources, Gaussian foregrounds and clusters).} The estimation of this bias term in the data measurement can be improved by replacing one of the simulated temperature field with data $\bar{D}$ to form a ``realization dependent $N^{(0)}$" \citep{namikawa13}:

\begin{align}
N_{L}^{(0),{\rm RD}}=\biggl\langle &C_{L}^{\widetilde\phi\widetilde\phi}[\widetilde\phi^{(1)}(\bar D\bar S_{i}^{\phi_{1}})\widetilde\phi^{(2)}(\bar D \bar S_{i}^{\phi_{1}})]\nonumber\\
+&C_{L}^{\widetilde\phi\widetilde\phi}[\widetilde\phi^{(1)}(\bar S_{i}^{\phi_{1}}\bar D)\widetilde\phi^{(2)}(\bar S_{i}^{\phi_{1}}\bar D)]\nonumber\\
+&C_{L}^{\widetilde\phi\widetilde\phi}[\widetilde\phi^{(1)}(\bar S_{i}^{\phi_{1}}\bar D)\widetilde\phi^{(2)}(\bar D\bar S_{i}^{\phi_{1}})]\nonumber\\
+&C_{L}^{\widetilde\phi\widetilde\phi}[\widetilde\phi^{(1)}(\bar D\bar S_{i}^{\phi_{1}})\widetilde\phi^{(2)}(\bar S_{i}^{\phi_{1}}\bar D)]\nonumber\\
-&C_{L}^{\widetilde\phi\widetilde\phi}[\widetilde\phi^{(1)}(\bar S_{i}^{\phi_{1}}\bar S_{j}^{\phi_{2}}) \widetilde\phi^{(2)}(\bar S_{i}^{\phi_{1}}\bar S_{j}^{\phi_{2}})]\nonumber \\
 -&C_{L}^{\widetilde\phi\widetilde\phi}[\widetilde\phi^{(1)}(\bar S_{i}^{\phi_{1}}\bar S_{j}^{\phi_{2}}) \widetilde\phi^{(2)}(\bar S_{j}^{\phi_{2}}\bar S_{i}^{\phi_{1}})]\biggl\rangle_{i,j}.
\end{align}

The $N^{(1)}$ bias arises from the coupling between the CMB and the lensing potential and can be obtained numerically by using combinations of simulation maps with the same lensing potential but different CMB realizations:
\begin{align}
N_{L}^{(1)}=\biggl\langle &C_{L}^{\widetilde\phi\widetilde\phi}[\widetilde\phi^{(1)}({\bar S}_{i}^{\phi_{1}}{\bar S}_{j}^{\phi_{1}})\widetilde\phi^{(2)}({\bar S}_{i}^{\phi_{1}}{\bar S}_{j}^{\phi_{1}})]\nonumber \\
+&C_{L}^{\widetilde\phi\widetilde\phi}[\widetilde\phi^{(1)}({\bar S}_{i}^{\phi_{1}}{\bar S}_{j}^{\phi_{1}})\widetilde\phi^{(2)}({\bar S}_{j}^{\phi_{1}}{\bar S}_{i}^{\phi_{1}})]\nonumber \\
-& C_{L}^{\widetilde\phi\widetilde\phi}[\widetilde\phi^{(1)}({\bar S}_{i}^{\phi_{1}}{\bar S}_{j}^{\phi_{2}})\widetilde\phi^{(2)}({\bar S}_{i}^{\phi_{1}}{\bar S}_{j}^{\phi_{2}})]\nonumber \\
-& C_{L}^{\widetilde\phi\widetilde\phi}[\widetilde\phi^{(1)}({\bar S}_{i}^{\phi_{1}}{\bar S}_{j}^{\phi_{2}})\widetilde\phi^{(2)}({\bar S}_{j}^{\phi_{2}}{\bar S}_{i}^{\phi_{1}})]\biggl\rangle_{i,j}.
\end{align}
To accelerate the calculation for the $N^{(1)}$ bias, we use simulations that contain the lensed CMB only.
\subsubsection{Foreground Biases}
Temperature-based lensing reconstruction is fractionally biased due to correlations with foreground components. We therefore consider the fractional lensing biases due to tSZ-4 point, CIB-4 point, ${\rm tSZ}^{2}$-$\phi$ and  ${\rm CIB}^{2}$-$\phi$ correlations as shown. We interpolate the results presented in \citealt{vanengelen14} using the four components as a function of $L$, and apply it to the fiducial model when calculating the best-fit amplitude.

\begin{figure*}
\begin{center}
\includegraphics[width=0.9\textwidth]{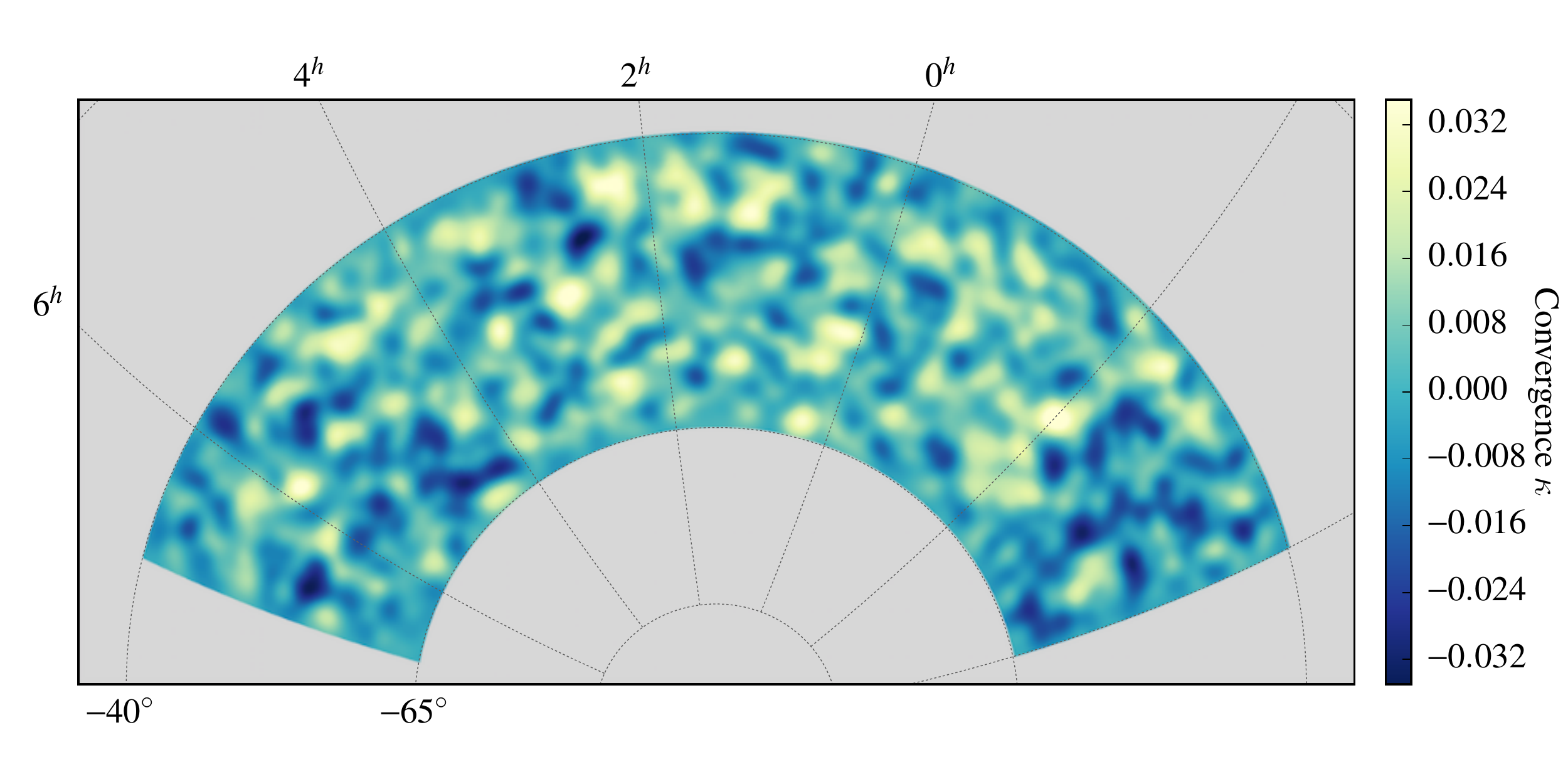}
\caption{The reconstructed lensing map on a zenithal equal-area projection. The map has been smoothed
with a Gaussian kernel with FWHM = 2 degrees.}
\label{fig:maps}
\end{center}
\end{figure*}

\begin{figure*}
\begin{center}
\includegraphics[width=1.0\textwidth]{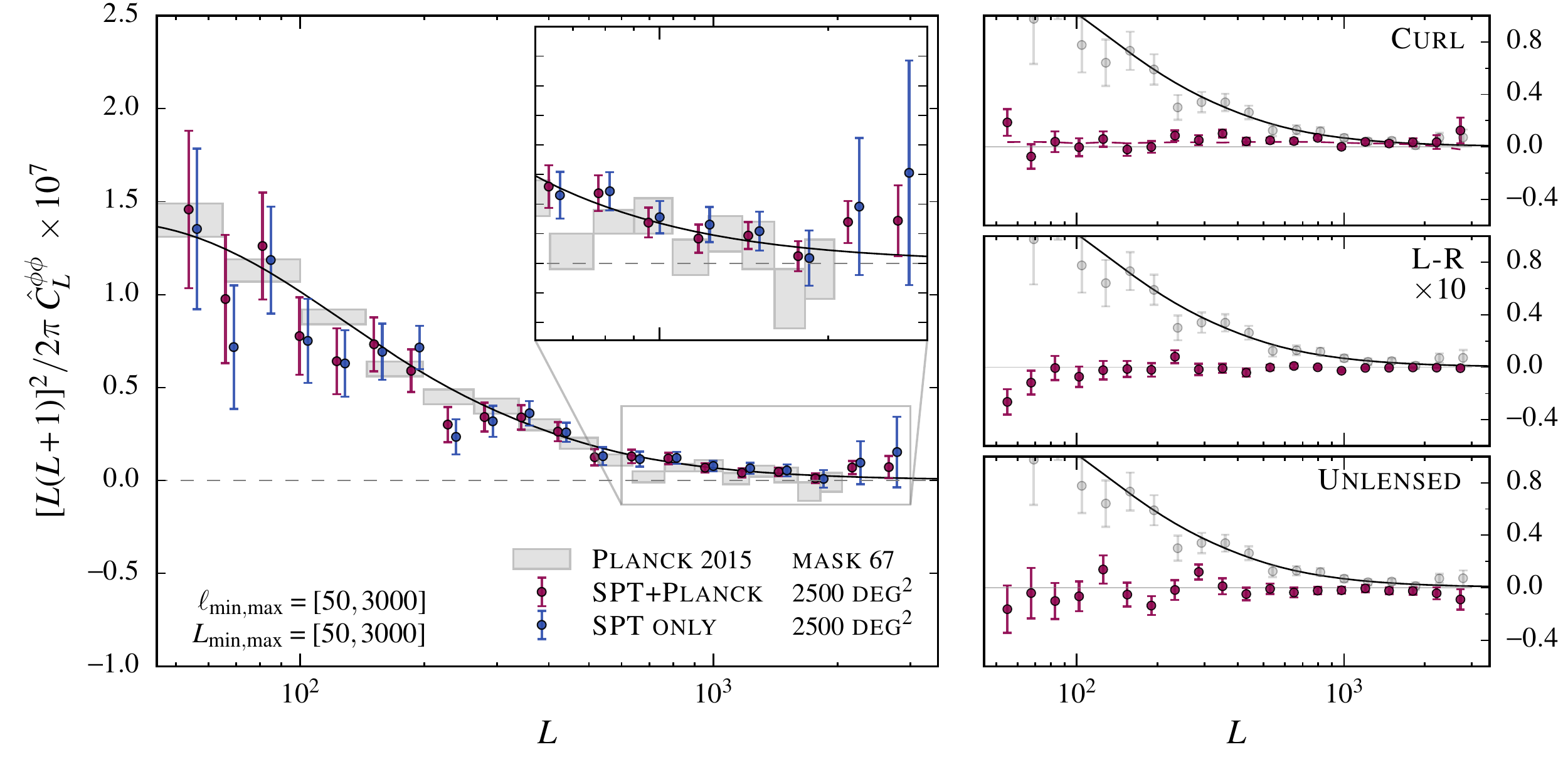}
\caption{Plot showing the consistency between the measured lensing auto-spectrum and the fiducial spectrum \emph{without the fractional lensing bias applied} (discussed in Section \ref{sec:clpp}), as well as the consistency of the curl, unlensed and L-R spectrum with respect to null (discussed in Section \ref{sec:nulltests}). {\bf Left:} \clpp auto-spectrum for SPT+\emph{Planck} (violet), SPT-only (blue),  and \emph{Planck} only using $\sim$67\% of the sky (gray boxes). {\xx The solid line is the fiducial $C_{L}^{\phi\phi}$ spectrum using a spatially flat $\Lambda$CDM \emph{Planck} 2015 cosmology}. A zoom-in of the high-$L$ region is presented to highlight the consistency between our measurement with our fiducial theory for $L<2000$ as well as the rise at $L>2000$, which is a possible indication of foreground contamination. {\bf Upper right:} The curl spectrum \clxx  calculated from the map. The solid violet line represents the mean of the simulation realizations, which is used to calculate the $\chi^{2}$ and PTE. {\bf Center right:} \clpp spectrum from one L-R realization with the amplitude multiplied by a factor of 10. {\bf Lower right:} \clpp spectrum for one unlensed realization. In each of 
the right panels, the fiducial result is shown in gray, and theory curve in black.}
\label{fig:clpp_main_null}
\end{center}
\begin{center}
\includegraphics[width=1.0\textwidth]{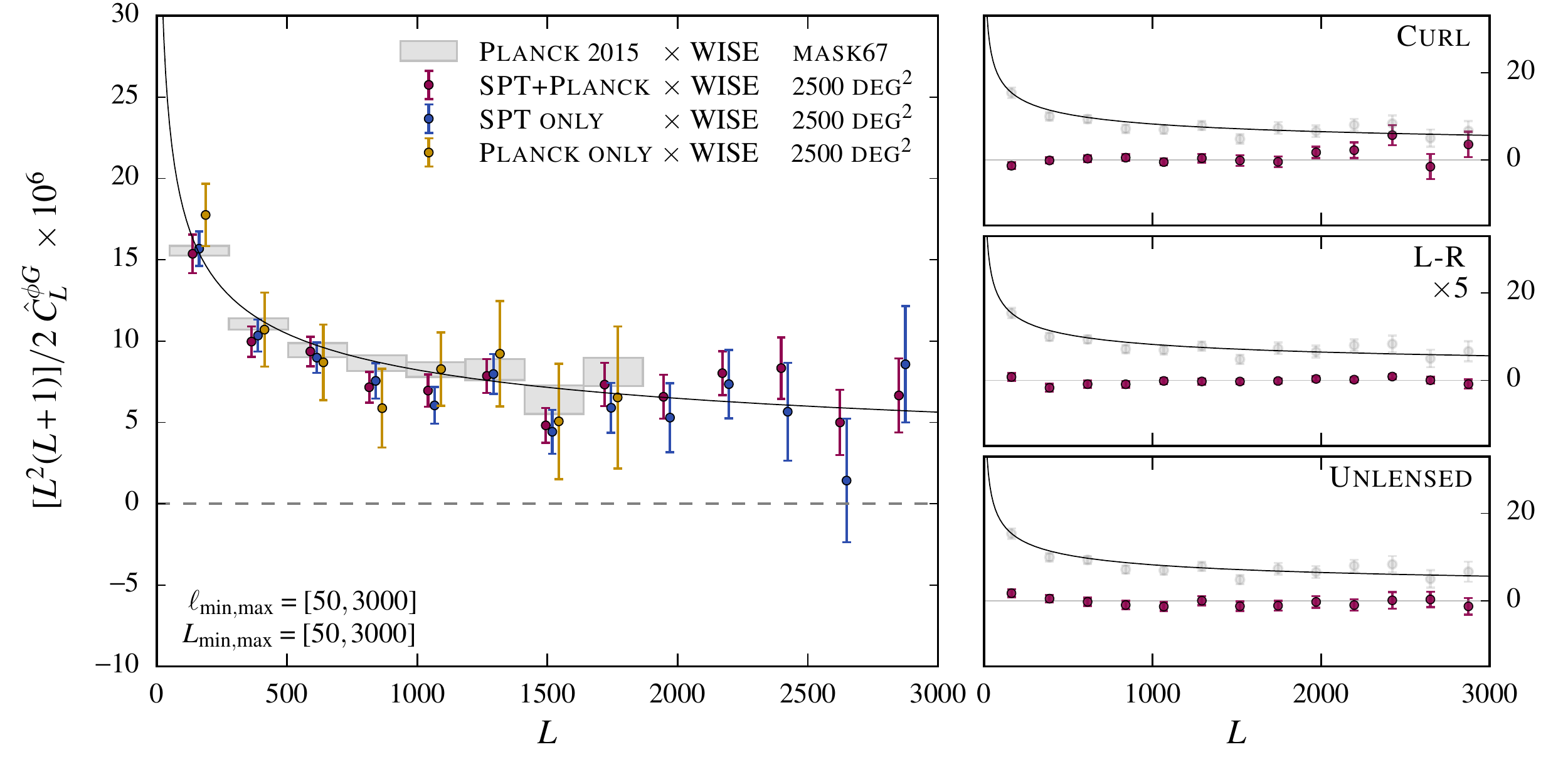}
\caption{A plot showing the consistency between SPT+\emph{Planck} and \emph{Planck}-only lensing, using the cross-correlation with \emph{WISE} galaxies (discussed in Section \ref{sec:cross}). Also shown are the cross-correlations between \emph{WISE} and curl, unlensed and L-R maps (discussed in Section \ref{sec:nulltests}).
 {\bf Left:} Cross-correlation between \emph{WISE} and: SPT+\emph{Planck} lensing map over $2500\ \sqdeg$ (violet), SPT only over $2500\ \sqdeg$ (blue), \emph{Planck} 2015 over the $2500\ \sqdeg$ (gold), and \emph{Planck} 2015 over the full sky (gray boxes). {\bf Right:} Cross-correlation of the galaxy sample with: {\bf upper right:} the data curl-mode map,  {\bf center right:}  a single realization of a noise-only reconstructed map,  {\bf lower right:} a single realization of an unlensed map. In each of the panels, a power-law fit to the \emph{Planck} result is shown as a solid line, and the fiducial result is shown as gray points in the three panels on the right.}
\label{fig:clkg_main_null2}
\end{center}
\end{figure*}

\section{Results}\label{sec:results}

\begin{figure}
\begin{center}
\includegraphics[width=0.50\textwidth]{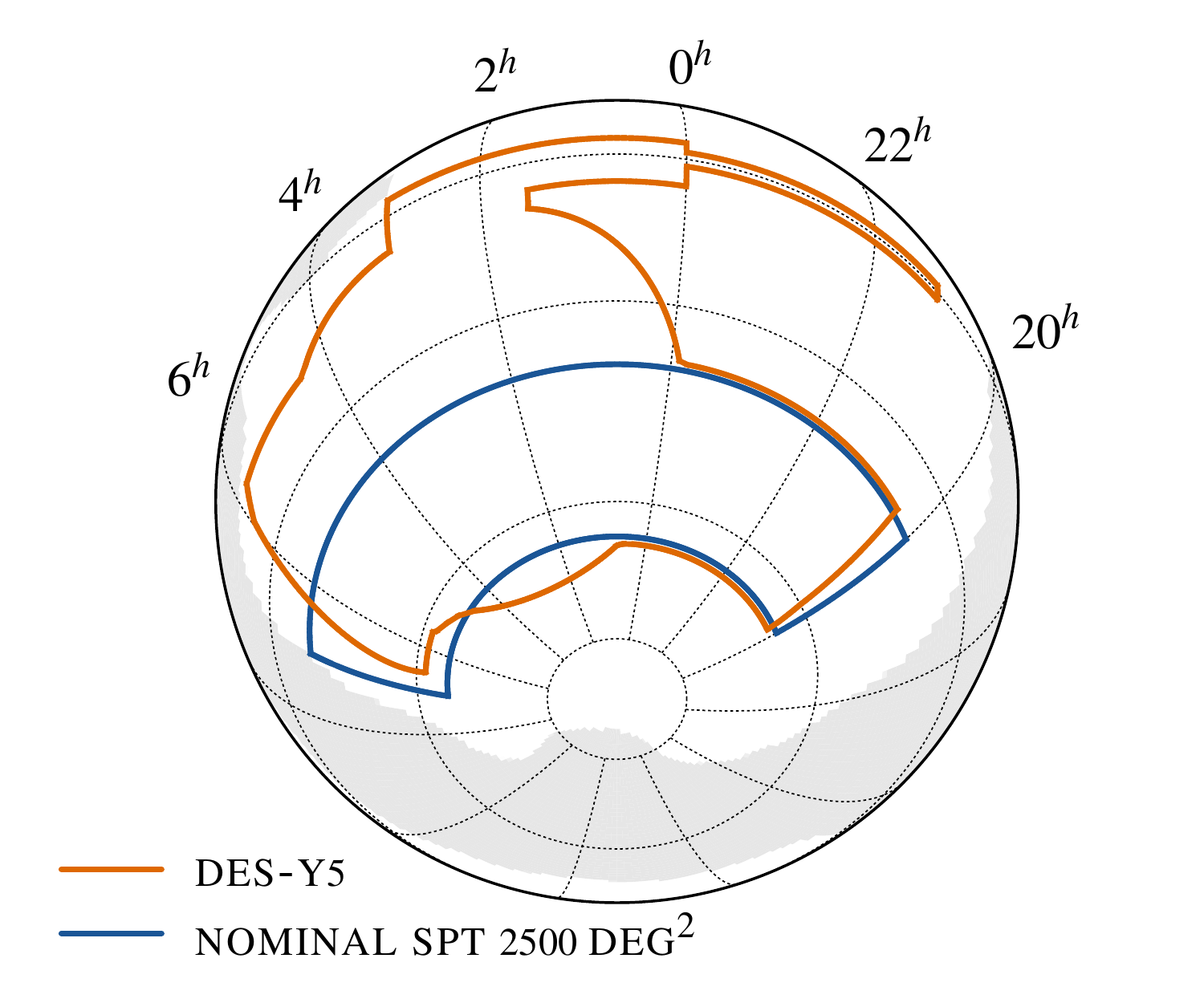}
\caption{{\bf Blue:} SPT nominal $2500\ {\rm deg}^{2}$ footprint, {\bf Orange:} DES-Y5 footprint, the expected coverage of DES after 5-years (2013-2018) of the survey. The \emph{Planck} galactic emission mask with $70\%$ coverage, based on 353 GHz emission is shown in light gray.}
\label{fig:footprint}
\end{center}
\end{figure}

{The SPT+\emph{Planck} $\hat \phi$ map reconstructed using the procedures outlined in Section \ref{sec:methods} is shown in Figure \ref{fig:maps} using a zenithal equal-area projection.  As shown in Figure \ref{fig:footprint}, this map has nearly full overlap with the final DES footprint. In this section, we 
validate the map with three calculations:
the auto-spectrum of the reconstructed $\hat \phi $ map and the cross-spectrum between the $\hat \phi $ map and two tracers of density fluctuations: \emph{WISE} galaxies and an estimate of the CIB from \emph{Planck} data.}

\subsection{$\hat C_{L}^{\phi\phi}$  Auto-spectrum}\label{sec:clpp}

{\xx In Figure \ref{fig:clpp_main_null}, we show the auto-spectrum of two versions of the reconstructed $\hat \phi$ map: one using SPT+\emph{Planck} and one with SPT data only. We also plot the $\hat \phi \hat \phi$ power spectrum from \emph{Planck} 2015 \citep{planck15-15}.} The auto-spectra for SPT+\emph{Planck} and SPT-only data in the range $50<L<3000$ are binned logarithmically using 20 bins, and the variance is calculated using the 198 simulation realizations. It should be noted that this is a slightly different binning {\xx scheme than employed} in \emph{Planck} 2015 and Simard et al. (in preparation), and that the points above $L>2000$ are likely to be affected by non-Gaussian foreground sources such as the CIB and tSZ from galaxies and low-mass galaxy clusters. The full $L$ range up to $L=3000$ is shown here for completeness to illustrate the raw spectrum from the lensing map itself. The ratio of the mean-field power and input spectrum is approximately unity at $L=50$. To ensure that the mean-field is not affecting our analyses, no modes below $L=50$ are considered.

We assign the name ``baseline" to the sample with  $\ell_{\rm max}=3000$, $\ell_{\rm min}=100$, $(\ell,m)$ cut $=[2000,250]$, clusters with $\xi>6$ masked, and point sources with $F_{150}>6.4\ {\rm mJy}$ masked. The ``SPT-only" sample is similar, but constructed bypassing the combining step and using $\ell_{\rm min}=550$.

We compare the SPT+\emph{Planck}, SPT-only and \emph{Planck} lensing auto-spectrum amplitudes relative to our fiducial model assuming diagonal covariance over the range $50<L<3000$. For {\xx the baseline sample of} SPT$+\emph{Planck}$, we obtain a best-fit amplitude of $\eta^{\phi\phi}=\hat C_{L}^{\phi\phi}/\hat C_{L,{\rm fid}}^{\phi\phi}=0.92^{+0.06}_{-0.06}$ with $\chi^{2}/\nu=12.2/19$. After removing the fractional lensing biases from foregrounds, we obtain $\eta^{\phi\phi}=\hat C_{L}^{\phi\phi}/\hat C_{L,{\rm fid}}^{\phi\phi}=0.95^{+0.06}_{-0.06}({\rm Stat.}) ^{+0.01} _{-0.01}({\rm Sys.})$ with $\chi^{2}/\nu=12.1/19$, where the goodness-of-fit is calculated using the statistical uncertainty only. Using the variance of unlensed simulations, we reject the null hypothesis of no lensing at $\sim 24\sigma$.

For SPT-only, we obtain $\eta^{\phi\phi}=0.91^{+0.06}_{-0.06}$ with $\chi^{2}/\nu=16.3/19$ when foreground biases are ignored and $\eta^{\phi\phi}=0.94^{+0.06}_{-0.06} ({\rm Stat.})\! ^{+0.01} _{-0.01}({\rm Sys.})$ with $\chi^{2}/\nu$ of $16.2/19$, when foreground biases are considered. In comparison, we obtain a best fit amplitude of $\eta^{\phi\phi}=0.98^{+0.02}_{-0.02}$ with $\chi^{2}/\nu$ of $25.1/18$ when \emph{Planck} band powers over $\sim67\% $ of the sky \citep{planck15-15} are fit to our fiducial model. The \emph{Planck} lensing map is less affected by foreground biases since it utilizes polarization in addition to temperature in the lensing reconstruction.

We find that the SPT+\emph{Planck} and SPT-only measurements are consistent with each other and with \emph{Planck} over $\sim67\%$ of the sky to within $0.5\sigma$. All the results reported here are summarized in Table \ref{table:results}.

\subsection{$\hat C_{L}^{\phi G}$  Cross-spectrum}\label{sec:cross}

An important purpose of this lensing map is for cross-correlation with external data sets. We calculate the cross-spectrum with the publicly available\footnote{\url{http://wise2.ipac.caltech.edu/docs/release/allsky/}} all-sky \emph{WISE} catalogue \citep{wright10}. This \emph{WISE} survey
mapped the sky at four wavelengths 3.4, 4.6, 12, and 22 $\mu m$ ($W1, W2, W3, W4$) with an angular resolution of $6.1, 6.4, 6.5,$ and $12.0$ arcseconds, respectively. We make one single cut in magnitude  $15<W1<17$ and remove all the flagged sources. The sample contains $2\times10^{8}$ sources in total using the mask employed in the $Planck$ lensing analysis, and $2\times10^{7}$ in sources in the nominal SPT region.
We make no attempt in estimating the redshift distribution of the galaxies, and hence cannot make a theoretical prediction of the cross-correlation amplitude. Instead, the lensing maps reconstructed using various $\ell_{\rm min,max}$, $(\ell,m)$ cuts, masking, and calibrations are cross-correlated with the galaxies to probe the sensitivity of the reconstructed lensing map to these variations.

Starting with the \emph{WISE} galaxy catalogue, we first project all the galaxies onto a {\tt HEALPix} map of {\tt $N_{\rm side}=2048$}, apply a simple binary mask (value=1 if there is at least one galaxy in the pixel, otherwise 0), and compute the mean number of galaxies $\langle n\rangle$.
Using this, the overdensity map is calculated:
\begin{equation}
\delta=\frac{n-\langle n \rangle}{\langle n \rangle},
\end{equation}
and the cross-spectrum is obtained by correlating this map with the lensing map using {\tt PolSpice}\footnote{\url{http://www2.iap.fr/users/hivon/software/PolSpice}}\citep{szapudi01,chon04}.

We derive the uncertainties by cross-correlating the \emph{WISE} galaxy density map with all the 198 simulated $\hat\phi$ maps and computing the variance for each bin. This method neglects the common sample variance between $\phi$ and the galaxies $G$. To assess the importance of this term, we compare this with errors obtained using the ``block jackknife" method (where the variance is computed by masking various ``blocks" of the sky area used in the analysis) with 128 equal area patches. We acquire similar results from this method and conclude that the original estimate is adequate.

Cross-spectra between \emph{WISE} galaxy density and various CMB-derived $\hat \phi$
are shown in Figure \ref{fig:clkg_main_null2}. The CMB lensing maps used are: SPT$+Planck$, SPT-only, \emph{Planck}-only over 2500 deg$^2$, and \emph{Planck}-only over 67\% of the sky.
We additionally sketch a power-law of the form $p_{L}=a(L/L_{0})^{-b}$, with parameters $a=2.15 \times 10^{-8}$, $b=1.35$, $L_{0}=490$, which are obtained by performing a least-squares fit to the cross-spectrum between full-sky \emph{Planck} and \emph{WISE} in the range $50<L<1864$. We then fit this power-law with an amplitude parameter $\eta^{\phi G}=C_{L}^{\phi G}/p_{L}$ to other cross-spectra. We obtain best-fit amplitudes of $\eta^{\phi G}=0.94^{+0.04}_{-0.04}$ for SPT+$Planck$,  $\eta^{\phi G}=0.93^{+0.04}_{-0.04}$ for SPT-only, $\eta^{\phi G}=1.00^{+0.02}_{-0.01}$ for $Planck$-only over $\sim67\%$ of the sky, and $\eta^{\phi G}=1.02^{+0.08}_{-0.08}$ for $Planck$-only over $2500\ {\rm deg}^{2}$.
Similar to the $\hat C_{L}^{\phi\phi}$ auto-spectrum, instead of focusing the discussion on the physical interpretations of the amplitude, which is dependent on factors such a photometric redshift uncertainties, type of galaxies considered, and the cosmological model used, we focus on the sensitivity of the cross-spectrum to small variations in the reconstruction pipeline.

\subsection{Cross-Correlation with CIB}\label{sec:sys_cib}
\begin{figure}
\begin{center}
\vspace{0.55cm}\includegraphics[width=0.5\textwidth]{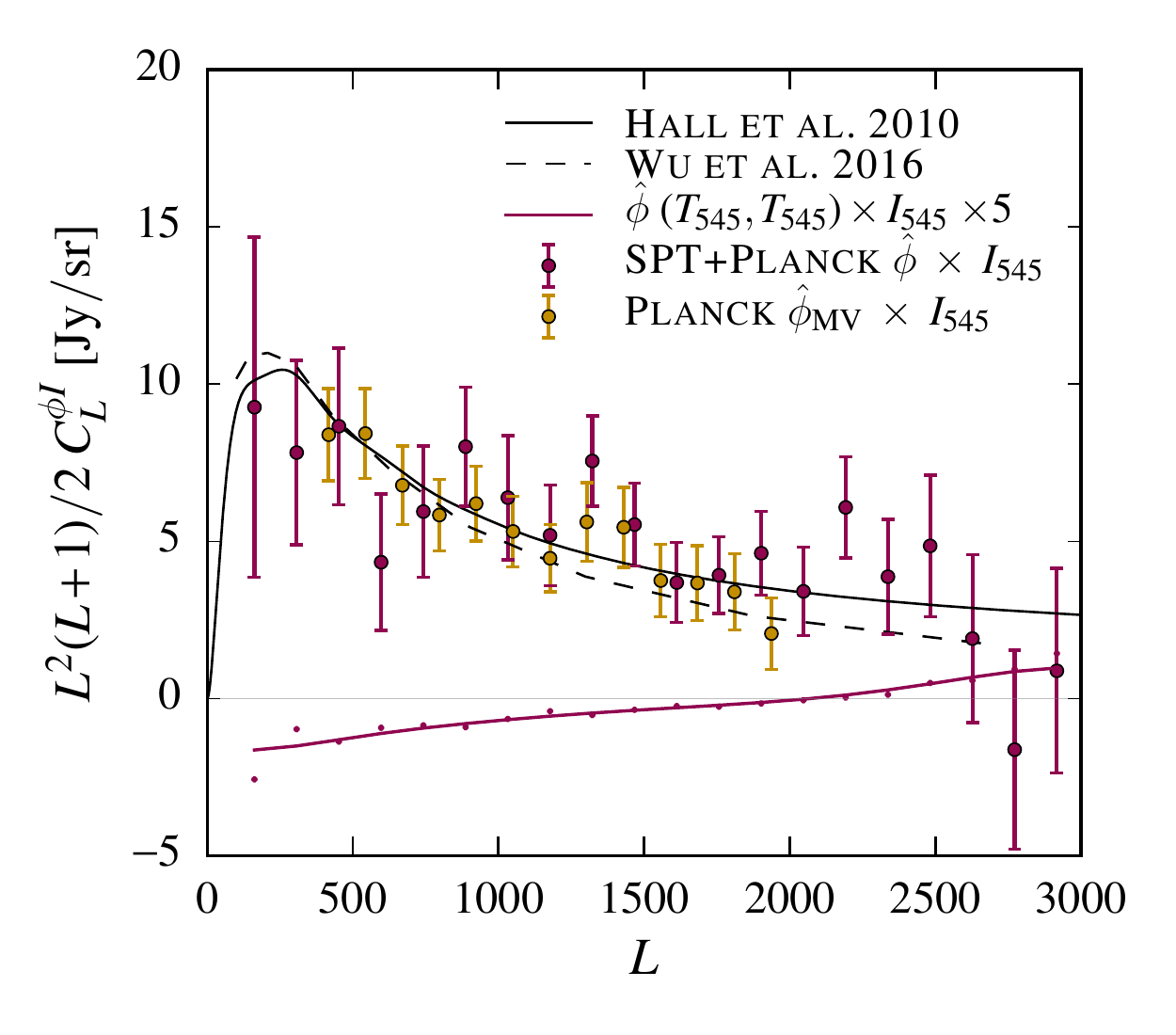}
\caption{{\xx Cross-correlation between \emph{Planck} 545 GHz channel and SPT+\emph{Planck} (violet), and results from \citealt{planck13-18} (gold). The solid violet line corresponds to the $ \hat\phi(T_{545},T_{545})\times I_{545}$ bispectrum calculated using the GNILC 545 GHz and \emph{Planck} 545 GHz maps, which contributes to the bias which affects the auto-spectrum. The amplitude of this spectrum is multiplied by a factor of 5 to highlight the characteristics as a function of $L$. The theoretical prediction by Wu et al. was obtained through private communications.}}
\label{fig:clkI}
\end{center}
\end{figure}

{\xx We also calculate the cross-correlation between the SPT+\emph{Planck} lensing map and the 545 GHz channel from \emph{Planck}\footnote{\url{HFI\_SkyMap\_545\_2048\_R2.02\_full.fits}}, which traces fluctuations in the CIB. The result is shown in Figure \ref{fig:clkI}, and the same measurement made by \emph{Planck} \citep{planck13-18} is also presented as a reference.
We observe a strong correlation between $\hat \phi$ and the 545 GHz map that is consistent with a theoretical model constructed using a modified black body and employing a single spectral energy distribution model as demonstrated in \citep{planck15-15}.

The SPT 150 GHz map and the \emph{Planck} 143 GHz map contain some emission from the CIB,
and leakage of this signal into the lensing map will bias the cross-correlation with the 545 GHz map.
To estimate the level of this bias,
we calculate the $\hat\phi(T_{545},T_{545})\times T_{545}$ bispectrum and scale the result
to approximate $\hat\phi(T_{150,\mathrm{CIB}},T_{150,\mathrm{CIB}})\times T_{545}$. Specifically, we use the GNILC component-separated CIB at 545 GHz\footnote{\url{COM\_CompMap\_CIB-GNILC-F545\_2048\_R2.00.fits}, publicly available at the \emph{Planck} legacy archive}, with the scaling to 150 GHz determined by taking the cross-correlation between 150 and 545 GHz maps.
The $\hat \phi$ component is calculated using the same masking and filtering as done in the making of $\hat\phi({T_{150},T_{150}})$, and this is cross-correlated with the 545 GHz channel. The bispectrum measured here is a result of CIB leaking into the 150 GHz channel.
We find the bias from the CIB in the 150 GHz / 143 GHz maps to be negative for angular multipoles $L<2000$ and positive for $L>2000$, with a fractional amplitude ranging from 2\% to 10\% in the range $50<L<3000$.}

\begin{deluxetable}{lcccc}
\tablewidth{0pt}
\tablecaption{Table summarizing the fits to fiducial theory without foreground biases considered}
\tablehead{
\colhead{Results} & \colhead{$\eta^{\phi\phi}$}          & \colhead{$\chi^{2}$\ (PTE)}  & \colhead{$\eta^{\phi G}$} & \colhead{$\chi^{2}$\ (PTE)}
}
\startdata
Baseline & $0.92^{+0.06}_{-0.06}$ & 12.2\ (0.88) & $0.94^{+0.04}_{-0.04}$ & 12.0\ (0.45) \\
SPT-only & $0.91^{+0.06}_{-0.06}$ & 16.3\ (0.63) & $0.93^{+0.04}_{-0.04}$ & 9.6\ (0.65) \\
\emph{Planck}-full & $0.98^{+0.02}_{-0.02}$ & 25.1\ (0.12) & $1.00^{+0.02}_{-0.01}$ & 6.1\ (0.53)\\
(spt patch) &  &  & $1.02^{+0.08}_{-0.08}$ & 3.8\ (0.80) \\ \hline\\[-0.15cm]

Curl & & 23.4\ (0.22) &  & 15.6\ (0.27)\\
L-R & & 28.6\ (0.10) &   & 11.1\ (0.60)\\
Unlensed & & 18.9\ (0.53) &  & 10.3\ (0.67)
\enddata
\label{table:results}
\end{deluxetable}\vspace{0.5cm}

\begin{deluxetable}{lcc}
\tablewidth{0pt}
\tablecaption{Table summarizing the fits to fiducial theory with foreground biases considered}
\tablehead{
\colhead{Results} & \colhead{$\eta^{\phi\phi}$}          & \colhead{$\chi^{2}$\ (PTE)}
}
\startdata
Baseline & $0.95^{+0.06}_{-0.06}({\rm stat.}) ^{+0.01}_{-0.01}({\rm sys.})$ & 12.1\ (0.88) \\
SPT-only & $0.94^{+0.06}_{-0.06}({\rm stat.}) ^{+0.01}_{-0.01}({\rm sys.})$ & 16.2\ (0.64) 
\enddata
\label{table:results}
\end{deluxetable}

\subsubsection{Gains from Adding Planck}

As shown in Figures \ref{fig:clpp_main_null} and \ref{fig:clkg_main_null2}, the signal in the combined 2500 deg$^2$ map comes mostly from the SPT data. Nonetheless, the addition of \emph{Planck} reduces the scatter for modes $L>1500$. In particular, the scatter in the cross-correlation for the angular bin $2762<L<3000$ is reduced by a factor of $\sim 2$. Characterizing the lensing map at high $L$ is important, especially for cross-correlation studies, since we could potentially probe astrophysical effects at these scales. The improvement is the result of additional mode pairs in the $\hat \phi$ reconstruction process;  for a particular lensing mode of interest $L$, the temperature mode pairs $\ell_{1}, \ell_{2}$ that are relevant satisfy the relationship $L=\ell_{1}+\ell_{2}$. Therefore, by introducing low-$\ell$ modes, the number of high-$\ell$+low-$\ell$ mode pairs increase, and the variance of a particular $L$ mode is reduced.

\section{Validations}\label{sec:sys}

The reconstructed $\hat \phi$ estimate from a map that contains no lensed CMB signal should be consistent with $\eta^{\phi\phi}=0$. This could potentially fail if the reconstruction process creates spurious temperature correlations that lead to false lensing signal. We therefore reconstruct maps that we expect to have no signal (curl, unlensed, and L-R maps) using the same pipeline and check that they are consistent with the null hypothesis.

Additionally, we also probe the robustness of the map by varying $\ell,m$ cuts, masking, calibration, beams and normalization method to verify that the map is insensitive to particular processing choices that we make.

\subsection{Null Tests}\label{sec:nulltests}

\subsubsection{Curl}
In estimating the lensing potential $\phi$, we have used the gradient component of the temperature field. It is instead possible to construct the curl-mode lensing field $\psi$, by replacing the weight function with:
\begin{align}
&W_{\ell_{1}\ell_{2}L}^{\psi}=-\mathstrut{\sqrt\frac{(2\ell_{1}+1)(2\ell_{2}+1)(2L+1)}{4\pi}}\nonumber\\
&\times C_{\ell_{1}}^{TT}\left(\frac{1-(-1)^{\ell_{1}+\ell_{2}+L}}{2}\right) \begin{pmatrix}
  \ell_{1} & \ell_{2} & L \\
  1 & 0 & -1\nonumber \\
 \end{pmatrix}\\
&\times \sqrt{L(L+1)\ell_{1}(\ell_{1}+1)}+(\ell_{1}\leftrightarrow\ell_{2}).
\end{align}
Sources of systematic contamination introduce non-Gaussianities in the reconstructed maps, which get decomposed into gradient and curl components.  Since we expect that physical phenomena produce negligible curl, a measurement of it is an indication of potential systematic bias in the gradient mode.
We reconstruct the $\hat\psi$ map in an identical manner subtracting off the $N_{L}^{(0),{\rm RD}}$ from the data and $N_{L}^{(0)}$ from the simulated realizations, but replacing the response function with that calculated for the $\hat\phi$. As noted in \cite{planck15-15,kesden03,vanengelen12,benoitlevy13}, the curl mode also includes a $N^{(1)}$ type bias.  However, in our analysis, instead of removing this term, we compare with the mean curl mode spectrum from the simulations, for which we obtain $\chi^{2}/\nu=23.4/19$, giving a PTE of 0.22 for the null hypothesis of no contamination.  From correlating the $\hat\psi$ map with the \emph{WISE} galaxy sample, we obtain a correlation that is consistent with respect to null with $\chi^{2}=15.6$ for 13 degrees of freedom giving a PTE of 0.27. The results are shown in the upper right panel of Figures \ref{fig:clpp_main_null} and \ref{fig:clkg_main_null2}.

\subsubsection{L-R Reconstruction}

Many potential sources of systematic instrumental contamination are coupled to the telescope scanning strategy.
We perform a null test for any contamination from systematic differences between opposite scan directions by reconstructing $\phi$ from null maps formed by differencing left-going (L) from right-going (R) scans.
We first combine the SPT L-R map with a noise realization of \emph{Planck} (since no L-R map exist for \emph{Planck}). We then pass this combined map through the lensing pipeline using the same filtering and response function as the regular case. From this, we obtain a $\chi^{2}/\nu=28.6/20$ for the auto, and $11.1/13$ for the cross-spectrum giving a PTE of  0.10 and 0.60 respectively. The results are shown in the center right panel of Figures \ref{fig:clpp_main_null} and \ref{fig:clkg_main_null2}.

\subsubsection{Unlensed Maps}

Lensing reconstruction relies on the non-Gaussian statistical properties that lensing imprints in the observed CMB. In the absence of lensing, the reconstructed $\phi$ map will be purely noise, and therefore, should be consistent with zero-signal. We simply test this by (i) replacing the lensed CMB with an unlensed CMB, (ii) producing both SPT and \emph{Planck} simulated skies, (iii) combining SPT and \emph{Planck}, (iv) running the lensing estimator in the same manner as a lensed realization, (v) and finally using the response function for the lensed case to produce a map. Since this makes use of simulations only, this is purely a test of the reconstruction pipeline.
We compute both the auto and cross-spectrum with \emph{WISE} using this map, and we see no evidence of inconsistency with respect to null. Measuring the $\chi^{2}/\nu$, we obtain 18.9/20 for auto, and 10.33/13 for the cross-spectrum giving a PTE of 0.53 and 0.67 respectively.
Maps reconstructed this way are used to calculate the significance of the no-lensing hypothesis, as well as estimating the lensing reconstruction noise, which is used for forecasting and covariance estimation. The results are shown in the lower right panel of Figures \ref{fig:clpp_main_null} and \ref{fig:clkg_main_null2}.

\subsection{Systematic Tests}
In this section, we modify certain aspects of the lensing reconstruction pipeline to test for systematic effects.  We quantify the effect by quoting the maximum deviation defined as ${\rm max}\{(\hat{C}_{L,{\rm modified}}-\hat{C}_{L,{\rm baseline}})/\sigma(\hat{C}_{L,{\rm baseline}})\}$ across all the bins in each systematics test relative to the statistical uncertainty. For auto-spectra, we additionally quantify the deviation of the systematically modified results from the baseline results by calculating the $\chi^2$ and corresponding PTE relative to zero, which are summarized in Table \ref{table:sys_table}. 
The same measurement is not carried out for cross-spectra since our method  of cross-correlating the galaxy map with different $\hat\phi$ realizations under-estimates the variance in ($\hat{C}_{L,{\rm sim}}^{\phi G}-\hat{C}_{L,{\rm sim,\ modified}}^{\phi G}$), which are the errorbars shown in Figure \ref{fig:clkg_sys_cross2}.
Nonetheless, the goal of this section is to illustrate that systematic variations lead to small changes in the resulting map, in comparison to the statistical uncertainty.

\subsubsection{$\ell_{\rm max}$, $\ell_{\rm min}$ Cut}\label{sec:sys_lmcut}
Contamination from point sources and {\xx the tSZ effect} is stronger at high $\ell$. {\xx Although including modes out to higher $\ell$ increases the number of modes one can use in the lensing reconstruction, bias due to the aforementioned contaminants also increases. We therefore apply a cut-off in the maximum $\ell$ used to minimize the bias.}  We set $\ell_{\rm max}$ for the baseline sample to $3000$, and make two alternatives cuts at $\ell_{\rm max}=2500$, and  $\ell_{\rm max}=3500$, to verify that the maps we obtain are not highly contaminated by foregrounds. As shown in Figure \ref{fig:clpp_sys_auto2}, we observe that changing $\ell_{\rm max}$ does effect the scatter, and the biggest change in any bin is seen when $\ell_{\rm max}$ is reduced to 2500 (maximum deviation of $1.4\sigma$ for the auto and 2.1$\sigma$ for the cross). When we vary $\ell_{\rm min}$ from 100 to 50,  we see negligible difference (maximum deviation of $0.027\sigma$ for $\hat{C}_{L}^{\phi\phi}$ and $0.10\sigma$ for $\hat{C}_{L}^{\phi G}$). The results are shown in the first panel of Figures \ref{fig:clpp_sys_auto2} and \ref{fig:clkg_sys_cross2}.

\subsubsection{$\ell,m$ Cuts}\label{sec:sys_lmcut}
High $\ell$, low $m$ modes of the combined map are dominated by noise since both SPT and \emph{Planck} are noisy for those modes. To remove the high noise modes, we apply cuts on the $(\ell,m)$ grid, and we test the sensitivity of the reconstructed $\phi$ map to this adopted cut. In calculating $\hat C_{L}^{\phi\phi}$, we calculate all the bias terms including $N_{L}^{(1)}$, using the same cuts, and obtain the response function in an identical fashion. We test three cuts which remove: (i) $\ell>2000$ \emph{and} $m<350$, (ii) $\ell>1200$ \emph{and} $m<350$ and (iii) $\ell>2200$ \emph{and} $m<150$. The comparison between the baseline sample and (i) demonstrates whether we are including excessive noise from SPT at low $m$. (ii) is a conservative cut in $\ell,m$, effectively removing noisy modes from \emph{both} SPT and \emph{Planck}. (iii) is the least conservative cut extending to higher $\ell$ and lower $m$. It should be noted that including slightly noisier temperature modes does not necessarily translate to noise bias since the filtering downweights these modes.
Sample (ii) shows the biggest deviation from the baseline sample with $0.82\sigma$ in $\hat{C}_{L}^{\phi \phi}$ and 0.83$\sigma$ in $\hat{C}_{L}^{\phi G}$. The results are shown in the second panel of Figures \ref{fig:clpp_sys_auto2} and \ref{fig:clkg_sys_cross2}.

\subsubsection{Cluster Masking}\label{sec:sys_masking}
One of the main concerns of temperature-based single-frequency lensing reconstruction is the contamination from the tSZ effect produced by clusters and galaxies. $\hat \phi$ maps reconstructed using temperature maps that contain tSZ power will be biased. The resulting bias in $\hat C_{L}^{\phi\phi}$ will include terms proportional to the tSZ 4-point function $\phi(T_{\rm tSZ}T_{\rm tSZ})\times\phi(T_{\rm tSZ}T_{\rm tSZ})$ and the $\phi$-tSZ correlation $\phi(T_{\rm CMB}T_{\rm CMB})\times\phi(T_{\rm tSZ}T_{\rm tSZ})$ {\xx \citep{vanengelen14}}. This bias will also result in a $\phi(T_{\rm tSZ}T_{\rm tSZ})\times G$ bispectrum when calculating cross-spectra with galaxies. The most dominant source of these biases are  massive clusters, which we mask in our analysis. We vary the masking radii and cluster selection to investigate the optimal masking to mitigate the contamination, while minimizing the sky area lost by the masking. 

\begin{figure*}
\begin{center}
\includegraphics[width=0.9\textwidth]{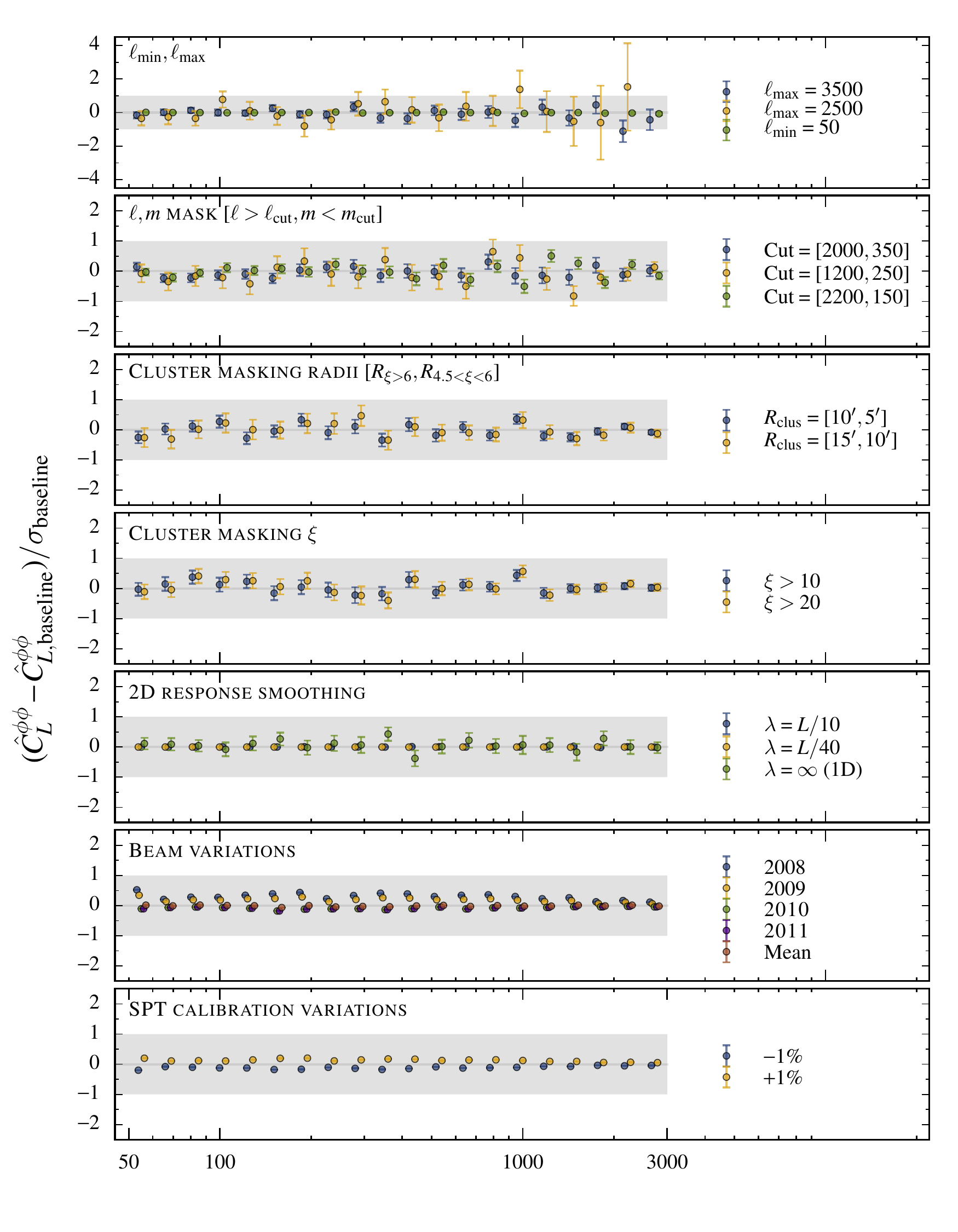}
\caption{Systematic tests for the lensing power spectrum $\hat{C}_{L}^{\phi\phi}$. Ratio of measured $\hat{C}_{L}^{\phi\phi}$ with variations made against input baseline $\hat{C}_{L}^{\phi\phi}$, where baseline $\hat{C}_{L}^{\phi\phi}$ is calculated using $\ell_{\rm min}=100, \ell_{\rm max}=3000, (\ell,m)$ cut at $[2000,250]$, $R_{\rm clus}=[5',5']$, $\sigma=L/20$ and beams appropriate for each SPT field (gray band). The error bars shown are the standard deviations of the \emph{change} in $\hat{C}_{L}^{\phi\phi}$ over a set of simulations.}
\label{fig:clpp_sys_auto2}
\end{center}
\end{figure*}

\begin{figure*}
\begin{center}
\includegraphics[width=0.9\textwidth]{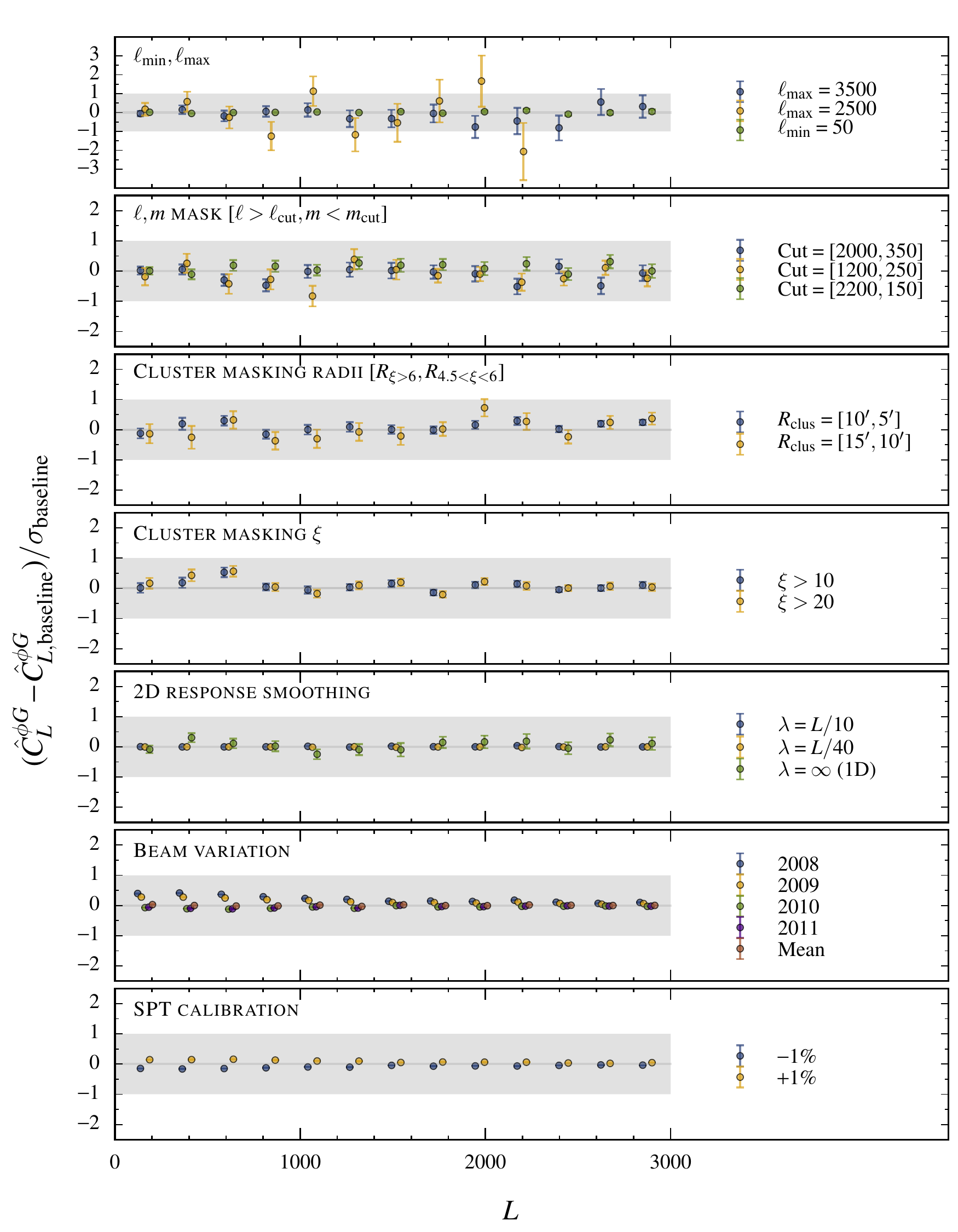}
\caption{Systematic tests for the lensing-galaxy cross-spectrum $\hat{C}_{L}^{\phi G}$. Ratio of measured $\hat{C}_{L}^{\phi G}$ with variations made against baseline $\hat{C}_{L}^{\phi G}$.}
\label{fig:clkg_sys_cross2}
\end{center}
\end{figure*}

 We make variations in the radius used to mask the locations of clusters in the reconstructed $\phi$ map. This is shown in the third panel of Figures \ref{fig:clpp_sys_auto2} and  \ref{fig:clkg_sys_cross2}. We tested using masks of larger radii for clusters with significance $\xi>6$ and $4.5<\xi<6$, using $R_{\xi>6}=10',15'$ and  $R_{4.6<\xi<6}=5',10'$. We found a maximum difference of only $\sim0.5$ and $\sim0.7\sigma$ discrepancies between our baseline auto and cross-spectrum respectively. The results are shown in the third panel of Figures \ref{fig:clpp_sys_auto2} and \ref{fig:clkg_sys_cross2}.

 Additionally, we mask clusters listed in \cite{bleem15b} with $\xi>6$ prior to running the quadratic estimator, and mask down further to  $\xi>4.5$ when calculating $\hat C_{L}^{\phi\phi}$ and $\hat C_{L}^{\phi G}$.  Tests in reconstructing $\phi$ maps with less strict cuts using $\xi=10,20$ are also made, and the results show that the $\hat C_{L}^{\phi\phi}$ amplitude for both cases are consistent with the $\xi>6$ cut sample with a maximum difference of $0.56\sigma$ for $\hat{C}_{L}^{\phi \phi}$ and $0.56\sigma$ for $\hat{C}_{L}^{\phi G}$. In calculating these spectra, a common mask that removes clusters above $\xi>4.5$ is applied to all maps. This test illustrates the amount of tSZ leakage during the reconstruction process. The results are shown in the fourth panel of Figures \ref{fig:clpp_sys_auto2} and \ref{fig:clkg_sys_cross2}.

\begin{deluxetable}{lcc}
\tablewidth{0pt}
\tablecaption{Table summarizing $\hat{C}_{L}^{\phi\phi}$ systematic test fits}
\tablecomments{$\chi^2$ and corresponding PTE relative to zero for the deviation of the systematically modified $\hat{C}_{L}^{\phi\phi}$ from the baseline $\hat{C}_{L}^{\phi\phi}$.}
\tablehead{
\colhead{Systematic change} & \colhead{$\chi^{2}/\nu$} & \colhead{PTE}}
\startdata
$\ell_{\rm max}=3500$    & 15.3/20 & 0.76 \\
$\ell_{\rm max}=2500$    & 10.5/19 & 0.94 \\
$\ell_{\rm min}=50  $    & 14.6/20 & 0.80 \\[0.2cm]

Cut=[2000,350]           & 15.6/20 & 0.74 \\
Cut=[1200,250]           & 19.6/20 & 0.48 \\
Cut=[2200,150]           & 31.5/20 & 0.05 \\[0.2cm]

$R_{\rm clus}=[10',5']$  & 27.8/20 & 0.11 \\
$R_{\rm clus}=[15',10']$ & 12.3/20 & 0.90 \\[0.2cm]

$\xi>10$                 & 16.1/20 & 0.71 \\
$\xi>20$                 & 22.8/20 & 0.30 \\[0.2cm]

$\lambda=L/10$           & 15.0/20 & 0.77 \\
$\lambda=L/40$           & 14.1/20 & 0.82 \\
$\lambda=\infty(1D)$     & 11.5/20 & 0.93 

\enddata
\label{table:sys_table}
\end{deluxetable}

\subsubsection{Response Function Smoothing}\label{sec:sys_respsmoothing}
Due to the large scatter at high $L$, the response function is smoothed to prevent the scatter appearing in \clpp and \clpg. The smoothed response function is shown in Figure \ref{fig:resp_2D} and the results of varying the smoothing length is shown in the fifth panel of Figures \ref{fig:clpp_sys_auto2} and \ref{fig:clkg_sys_cross2}. In both auto and cross-spectra, the variations show negligible differences, with maximum discrepancy of 0.3$\sigma$ when using a 1D response function. The results are shown in the fifth panel of Figures \ref{fig:clpp_sys_auto2} and \ref{fig:clkg_sys_cross2}.

\subsubsection{Beam Error}\label{sec:beam_err}

The various SPT fields were observed in different years (2008--2011), and instrumental changes to the receiver between observing years result in a slightly different beam for each year and, hence, each field. In the baseline analysis, we deconvolve each field with a specific year beam, and convolve with a common Gaussian beam of FWHM=$1.75'$. We also test the lensing reconstruction using (i) the four single year beams for all the fields and (ii) the average of the year beams for all the fields. This effectively probes the sensitivity of the resulting map to the uncertainty of the beam.

The effect of the beam is most prominently seen in $\hat{C}_{L}^{\phi G}$ with a maximum deviation of $0.41\sigma$ when assuming a 2008 beam. Deconvolving all the fields with a mean beam produces a maximum deviation of 0.021$\sigma$ and 0.029$\sigma$ for $\hat{C}_{L}^{\phi\phi}$ and $\hat{C}_{L}^{\phi G}$ respectively, suggesting that it is a good approximation of the baseline method. The results are shown in the sixth panel of Figures \ref{fig:clpp_sys_auto2} and \ref{fig:clkg_sys_cross2}.

\subsubsection{Calibration Error}\label{sec:sys_spt_calb}
The CMB power as measured by SPT is calibrated to align with the measurements made by \emph{Planck} in the same patch of sky to an accuracy better than 1\% \citep{hou17}. The results of varying this calibration parameter by $\pm 1$\% is shown in the bottom panel of Figures \ref{fig:clpp_sys_auto2} and \ref{fig:clkg_sys_cross2}. The resulting \clpp and \clpg vary by at most $0.20\sigma$ and $0.16\sigma$ respectively through this variation. The results are shown in the seventh panel of Figures \ref{fig:clpp_sys_auto2} and \ref{fig:clkg_sys_cross2}.

\section{Discussions}\label{sec:discussions}
This paper presents a map of the CMB lensing potential over $\sim$2500 square degrees of the sky, constructed from the optimally combined SPT\ 150\ GHz + \emph{Planck}\ 143\ GHz temperature map. The cosmological constraints from this data set will be published in a companion paper (Simard et al, in preparation).

The lensing map has an improved signal-to-noise ratio at all scales relative to \emph{Planck}-only over the $2500\ {\rm deg}^{2}$ patch, and it overlaps almost completely with the DES galaxy survey, making it a potentially powerful data set for cross-correlation studies.
The power of this lensing map comes primarily from the 2500-square-degree SPT-SZ survey data, with typical map noise of 18 $\mu$K-arcmin, but adding \emph{Planck}\ 143\ GHz data results in noticeable improvement in S/N, especially at high lensing multipole $L$. As shown in Figures \ref{fig:clpp_main_null} and \ref{fig:clkg_main_null2}, by filling in the SPT modes with low-$\ell$ modes from \emph{Planck}, moderate improvements can be observed at intermediate to high $L\ (1500<L<3000)$, which is due to the increased number of mode pairs that enter the reconstruction.
We compare the measured lensing amplitude with the amplitude expected from a fiducial $\Lambda {\rm CDM}$ \emph{Planck} 2015 best fit cosmology and obtain $\eta^{\phi\phi}=0.95_{-0.06}^{+0.06}({\rm Stat.})\! _{-0.01}^{+0.01}({\rm Sys.})$.
The total lensing S/N in the combined map is approximately $14 \sigma$, and this measurement rejects the no-lensing hypothesis at $\sim24\sigma$.

We perform several consistency checks, null tests, and systematic tests on the SPT + \emph{Planck} lensing map. We measure the auto-spectrum of the lensing map and the cross-spectrum of the lensing map with \emph{WISE} galaxy positions, and we find both results to be consistent with expectations from the fiducial cosmology, with \emph{Planck} full-sky results, and with SPT-only or \emph{Planck}-only results over the 2500-square-degree SPT-SZ region.
We perform null tests using the curl lensing estimator on the combined temperature map and by replacing the temperature map by unlensed CMB or L-R difference map, and we find these results to be consistent with noise.
Finally, we investigate potential systematics by recalculating the lensing auto-spectrum and lensing-\emph{WISE} cross-spectrum with certain parameters shifted and comparing to our baseline result. We vary several sets of parameters, including the $(\ell,m)$ range used in the lensing reconstruction, the way in which emissive sources and galaxy clusters are masked, and the way beam and calibration errors are treated. We find no evidence of systematic contamination from these tests.

With the SPT 2500\ {\rm deg}$^{2}$ field almost fully overlapping with DES, it is possible to perform cross-correlations with various mass tracers. {\xx A number of cross-correlations} between a lensing map produced from a small SPT patch and DES SV data, which covered $\sim 140\ {\rm deg}^{2}$  have been measured to date: cross-correlation with galaxy density \citep{giannantonio16}, galaxy shear \citep{kirk15}, and the ratio between galaxy-galaxy lensing and galaxy CMB lensing \citep{baxter16}. {\xx With future DES releases}, which will have more than {\rm $2000\ {\rm deg}^{2}$} overlap, these measurements are expected to improve significantly, allowing us to place tighter cosmological parameter constraints.

In this work, we only use temperature data to a maximum multipole of $\ell = 3000$ due to concerns of foreground contamination. Obtaining a reliable lensing map at high $\ell$ {\xx requires precise} knowledge of the foreground components that contribute to the small scale power.
It is, however, a challenging task to separate the contributions from various foreground components without using data from multiple frequencies. In this analysis, the noise levels of the 95 and 220 GHz channels of the SPT-SZ survey (40 $\mu$K-arcmin and 70 $\mu$K-arcmin respectively), prohibit us from removing foregrounds cleanly at the level required for lensing analysis. We look forward to do multi-frequency lensing analysis using SPT-3G \citep{benson14}, which will have multiple channels at lower noise levels than the SPT-SZ survey. Furthermore, since extra-galactic foreground contamination is a much smaller effect in polarization maps, we will be able to utilize information at higher $\ell$ compared to the temperature data\citep{osborne14}.

\section{Acknowledgments}
We thank Hao-Yi Wu and Olivier Dor\'{e} for the theoretical prediction for the cross-correlation between CIB and CMB lensing.
The South Pole Telescope program is supported by the National Science Foundation through grant PLR-1248097. Partial support is also provided by the NSF Physics Frontier Center grant PHY-0114422 to the Kavli Institute of Cosmological Physics at the University of Chicago, the Kavli Foundation, and the Gordon and Betty Moore Foundation through Grant GBMF\#947 to the University of Chicago. The McGill authors acknowledge funding from the Natural Sciences and Engineering Research Council of Canada, Canadian Institute for Advanced Research, and Canada Research Chairs program.
CR acknowledges support from a Australian Research Council Future Fellowship (FT150100074)
BB is supported by the Fermi Research Alliance, LLC under Contract No. De-AC02-07CH11359 with the United States Department of Energy.
This publication makes use of data products from the Wide-field Infrared Survey Explorer, which is a joint project of the University of California, Los Angeles, and the Jet Propulsion Laboratory/California Institute of Technology, funded by the National Aeronautics and Space Administration.
Argonne National Laboratory’s work was supported under U.S. Department of Energy contract DE-AC02-06CH11357.
Computations were made on the supercomputer Guillimin from McGill University, managed by Calcul Qu\'{e}bec and Compute Canada. The operation of this supercomputer is funded by the Canada Foundation for Innovation (CFI), the minist\`{e}re de l'\'{E}conomie, de la science et de l'innovation du Qu\'{e}bec (MESI) and the Fonds de recherche du Qu\'{e}bec - Nature et technologies (FRQ-NT).
This research is part of the Blue Waters sustained-petascale computing project, which is supported by the National Science Foundation (awards OCI-0725070 and ACI-1238993) and the state of Illinois. Blue Waters is a joint effort of the University of Illinois at Urbana-Champaign and its National Center for Supercomputing Applications.
This work used resources made available on the Jupiter cluster, a joint data-intensive computing project between the High Energy Physics Division and the Computing, Environment, and Life Sciences (CELS) Directorate at Argonne National Laboratory.
The results in this paper have been derived using the following packages: {\tt Astropy}, a community-developed core Python package for Astronomy \citep{astropy13}, {\tt CAMB} \citep{lewis00,howlett12}, {\tt HEALPix} \citep{gorski05}, {\tt IPython} \citep{ipython}, {\tt LensPix} \citep{lewis05}, {\tt Matplotlib} \citep{hunter07}, {\tt NumPy} \& {\tt SciPy} \citep{numpy}, {\tt PolSpice} \citep{chon04} and {\tt quicklens} \citep{planck15-15}.
\newpage
\bibliographystyle{fapj}
\bibliography{spt}

\begin{thebibliography}{59}
\expandafter\ifx\csname natexlab\endcsname\relax\def\natexlab#1{#1}\fi

\bibitem[{{Abazajian} {et~al.}(2015){Abazajian}, {Arnold}, {Austermann},
  {Benson}, {Bischoff}, {Bock}, {Bond}, {Borrill}, {Calabrese}, {Carlstrom},
  {Carvalho}, {Chang}, {Chiang}, {Church}, {Cooray}, {Crawford}, {Dawson},
  {Das}, {Devlin}, {Dobbs}, {Dodelson}, {Dor{\'e}}, {Dunkley}, {Errard},
  {Fraisse}, {Gallicchio}, {Halverson}, {Hanany}, {Hildebrandt}, {Hincks},
  {Hlozek}, {Holder}, {Holzapfel}, {Honscheid}, {Hu}, {Hubmayr}, {Irwin},
  {Jones}, {Kamionkowski}, {Keating}, {Keisler}, {Knox}, {Komatsu}, {Kovac},
  {Kuo}, {Lawrence}, {Lee}, {Leitch}, {Linder}, {Lubin}, {McMahon}, {Miller},
  {Newburgh}, {Niemack}, {Nguyen}, {Nguyen}, {Page}, {Pryke}, {Reichardt},
  {Ruhl}, {Sehgal}, {Seljak}, {Sievers}, {Silverstein}, {Slosar}, {Smith},
  {Spergel}, {Staggs}, {Stark}, {Stompor}, {Vieregg}, {Wang}, {Watson},
  {Wollack}, {Wu}, {Yoon}, \& {Zahn}}]{abazajian15b}
{Abazajian}, K.~N., {et~al.} 2015, Astroparticle Physics, 63, 66

\bibitem[{{Astropy Collaboration} {et~al.}(2013){Astropy Collaboration},
  {Robitaille}, {Tollerud}, {Greenfield}, {Droettboom}, {Bray}, {Aldcroft},
  {Davis}, {Ginsburg}, {Price-Whelan}, {Kerzendorf}, {Conley}, {Crighton},
  {Barbary}, {Muna}, {Ferguson}, {Grollier}, {Parikh}, {Nair}, {Unther},
  {Deil}, {Woillez}, {Conseil}, {Kramer}, {Turner}, {Singer}, {Fox}, {Weaver},
  {Zabalza}, {Edwards}, {Azalee Bostroem}, {Burke}, {Casey}, {Crawford},
  {Dencheva}, {Ely}, {Jenness}, {Labrie}, {Lim}, {Pierfederici}, {Pontzen},
  {Ptak}, {Refsdal}, {Servillat}, \& {Streicher}}]{astropy13}
{Astropy Collaboration}, {et~al.} 2013, \aap, 558, A33

\bibitem[{{Baxter} {et~al.}(2016){Baxter}, {Clampitt}, {Giannantonio},
  {Dodelson}, {Jain}, {Huterer}, {Bleem}, {Crawford}, {Efstathiou}, {Fosalba},
  {Kirk}, {Kwan}, {S{\'a}nchez}, {Story}, {Troxel}, {Abbott}, {Abdalla},
  {Armstrong}, {Benoit-L{\'e}vy}, {Benson}, {Bernstein}, {Bernstein}, {Bertin},
  {Brooks}, {Carlstrom}, {Rosell}, {Carrasco Kind}, {Carretero}, {Chown},
  {Crocce}, {Cunha}, {da Costa}, {Desai}, {Diehl}, {Dietrich}, {Doel},
  {Evrard}, {Fausti Neto}, {Flaugher}, {Frieman}, {Gruen}, {Gruendl},
  {Gutierrez}, {de Haan}, {Holder}, {Honscheid}, {Hou}, {James}, {Kuehn},
  {Kuropatkin}, {Lima}, {March}, {Marshall}, {Martini}, {Melchior}, {Miller},
  {Miquel}, {Mohr}, {Nord}, {Omori}, {Plazas}, {Reichardt}, {Romer}, {Rykoff},
  {Sanchez}, {Sevilla-Noarbe}, {Sheldon}, {Smith}, {Soares-Santos}, {Sobreira},
  {Suchyta}, {Stark}, {Swanson}, {Tarle}, {Thomas}, {Walker}, \&
  {Wechsler}}]{baxter16}
{Baxter}, E., {et~al.} 2016, \mnras, 461, 4099

\bibitem[{{Bennett} {et~al.}(2003){Bennett}, {Halpern}, {Hinshaw}, {Jarosik},
  {Kogut}, {Limon}, {Meyer}, {Page}, {Spergel}, {Tucker}, {Wollack}, {Wright},
  {Barnes}, {Greason}, {Hill}, {Komatsu}, {Nolta}, {Odegard}, {Peiris},
  {Verde}, \& {Weiland}}]{bennett03a}
{Bennett}, C.~L., {et~al.} 2003, \apjs, 148, 1

\bibitem[{{Benoit-L{\'e}vy} {et~al.}(2013){Benoit-L{\'e}vy}, {D{\'e}chelette},
  {Benabed}, {Cardoso}, {Hanson}, \& {Prunet}}]{benoitlevy13}
{Benoit-L{\'e}vy}, A., {D{\'e}chelette}, T., {Benabed}, K., {Cardoso}, J.-F.,
  {Hanson}, D., \& {Prunet}, S. 2013, \aap, 555, A37

\bibitem[{{Benson} {et~al.}(2014){Benson}, {Ade}, {Ahmed}, {Allen}, {Arnold},
  {Austermann}, {Bender}, {Bleem}, {Carlstrom}, {Chang}, {Cho}, {Ciocys},
  {Cliche}, {Crawford}, {Cukierman}, {de Haan}, {Dobbs}, {Dutcher}, {Everett},
  {Gilbert}, {Halverson}, {Hanson}, {Harrington}, {Hattori}, {Henning},
  {Hilton}, {Holder}, {Holzapfel}, {Irwin}, {Keisler}, {Knox}, {Kubik}, {Kuo},
  {Lee}, {Leitch}, {Li}, {McDonald}, {Meyer}, {Montgomery}, {Myers}, {Natoli},
  {Nguyen}, {Novosad}, {Padin}, {Pan}, {Pearson}, {Reichardt}, {Ruhl},
  {Saliwanchik}, {Simard}, {Smecher}, {Sayre}, {Shirokoff}, {Stark}, {Story},
  {Suzuki}, {Thompson}, {Tucker}, {Vanderlinde}, {Vieira}, {Vikhlinin}, {Wang},
  {Yefremenko}, \& {Yoon}}]{benson14}
{Benson}, B.~A., {et~al.} 2014, in Society of Photo-Optical Instrumentation
  Engineers (SPIE) Conference Series, Vol. 9153, Society of Photo-Optical
  Instrumentation Engineers (SPIE) Conference Series

\bibitem[{{Bernardeau} {et~al.}(1997){Bernardeau}, {van Waerbeke}, \&
  {Mellier}}]{bernardeau97a}
{Bernardeau}, F., {van Waerbeke}, L., \& {Mellier}, Y. 1997, \aap, 322, 1

\bibitem[{{Bleem} {et~al.}(2015){Bleem}, {Stalder}, {de Haan}, {Aird}, {Allen},
  {Applegate}, {Ashby}, {Bautz}, {Bayliss}, {Benson}, {Bocquet}, {Brodwin},
  {Carlstrom}, {Chang}, {Chiu}, {Cho}, {Clocchiatti}, {Crawford}, {Crites},
  {Desai}, {Dietrich}, {Dobbs}, {Foley}, {Forman}, {George}, {Gladders},
  {Gonzalez}, {Halverson}, {Hennig}, {Hoekstra}, {Holder}, {Holzapfel},
  {Hrubes}, {Jones}, {Keisler}, {Knox}, {Lee}, {Leitch}, {Liu}, {Lueker},
  {Luong-Van}, {Mantz}, {Marrone}, {McDonald}, {McMahon}, {Meyer}, {Mocanu},
  {Mohr}, {Murray}, {Padin}, {Pryke}, {Reichardt}, {Rest}, {Ruel}, {Ruhl},
  {Saliwanchik}, {Saro}, {Sayre}, {Schaffer}, {Schrabback}, {Shirokoff},
  {Song}, {Spieler}, {Stanford}, {Staniszewski}, {Stark}, {Story}, {Stubbs},
  {Vanderlinde}, {Vieira}, {Vikhlinin}, {Williamson}, {Zahn}, \&
  {Zenteno}}]{bleem15b}
{Bleem}, L.~E., {et~al.} 2015, \apjs, 216, 27

\bibitem[{{Carlstrom} {et~al.}(2011){Carlstrom}, {Ade}, {Aird}, {Benson},
  {Bleem}, {Busetti}, {Chang}, {Chauvin}, {Cho}, {Crawford}, {Crites}, {Dobbs},
  {Halverson}, {Heimsath}, {Holzapfel}, {Hrubes}, {Joy}, {Keisler}, {Lanting},
  {Lee}, {Leitch}, {Leong}, {Lu}, {Lueker}, {Luong-van}, {McMahon}, {Mehl},
  {Meyer}, {Mohr}, {Montroy}, {Padin}, {Plagge}, {Pryke}, {Ruhl}, {Schaffer},
  {Schwan}, {Shirokoff}, {Spieler}, {Staniszewski}, {Stark}, {Tucker},
  {Vanderlinde}, {Vieira}, \& {Williamson}}]{carlstrom11}
{Carlstrom}, J.~E., {et~al.} 2011, \pasp, 123, 568

\bibitem[{{Cavaliere} \& {Fusco-Femiano}(1976)}]{cavaliere76}
{Cavaliere}, A., \& {Fusco-Femiano}, R. 1976, \aap, 49, 137

\bibitem[{{Chon} {et~al.}(2004){Chon}, {Challinor}, {Prunet}, {Hivon}, \&
  {Szapudi}}]{chon04}
{Chon}, G., {Challinor}, A., {Prunet}, S., {Hivon}, E., \& {Szapudi}, I. 2004,
  \mnras, 350, 914

\bibitem[{{Cooray} {et~al.}(2005){Cooray}, {Kamionkowski}, \&
  {Caldwell}}]{cooray05}
{Cooray}, A., {Kamionkowski}, M., \& {Caldwell}, R.~R. 2005, \prd, 71, 123527

\bibitem[{{Crawford} {et~al.}(2016){Crawford}, {Chown}, {Holder}, {Aird},
  {Benson}, {Bleem}, {Carlstrom}, {Chang}, {Cho}, {Crites}, {de Haan}, {Dobbs},
  {George}, {Halverson}, {Harrington}, {Holzapfel}, {Hou}, {Hrubes}, {Keisler},
  {Knox}, {Lee}, {Leitch}, {Luong-Van}, {Marrone}, {McMahon}, {Meyer},
  {Mocanu}, {Mohr}, {Natoli}, {Padin}, {Pryke}, {Reichardt}, {Ruhl}, {Sayre},
  {Schaffer}, {Shirokoff}, {Staniszewski}, {Stark}, {Story}, {Vanderlinde},
  {Vieira}, \& {Williamson}}]{crawford16}
{Crawford}, T.~M., {et~al.} 2016, \apjs, 227, 23

\bibitem[{{Das} {et~al.}(2011){Das}, {Sherwin}, {Aguirre}, {Appel}, {Bond},
  {Carvalho}, {Devlin}, {Dunkley}, {D{\"u}nner}, {Essinger-Hileman}, {Fowler},
  {Hajian}, {Halpern}, {Hasselfield}, {Hincks}, {Hlozek}, {Huffenberger},
  {Hughes}, {Irwin}, {Klein}, {Kosowsky}, {Lupton}, {Marriage}, {Marsden},
  {Menanteau}, {Moodley}, {Niemack}, {Nolta}, {Page}, {Parker}, {Reese},
  {Schmitt}, {Sehgal}, {Sievers}, {Spergel}, {Staggs}, {Swetz}, {Switzer},
  {Thornton}, {Visnjic}, \& {Wollack}}]{das11}
{Das}, S., {et~al.} 2011, Physical Review Letters, 107, 021301

\bibitem[{{De Zotti} {et~al.}(2005){De Zotti}, {Ricci}, {Mesa}, {Silva},
  {Mazzotta}, {Toffolatti}, \& {Gonz{\'a}lez-Nuevo}}]{dezotti05}
{De Zotti}, G., {Ricci}, R., {Mesa}, D., {Silva}, L., {Mazzotta}, P.,
  {Toffolatti}, L., \& {Gonz{\'a}lez-Nuevo}, J. 2005, \aap, 431, 893

\bibitem[{{George} {et~al.}(2015){George}, {Reichardt}, {Aird}, {Benson},
  {Bleem}, {Carlstrom}, {Chang}, {Cho}, {Crawford}, {Crites}, {de Haan},
  {Dobbs}, {Dudley}, {Halverson}, {Harrington}, {Holder}, {Holzapfel}, {Hou},
  {Hrubes}, {Keisler}, {Knox}, {Lee}, {Leitch}, {Lueker}, {Luong-Van},
  {McMahon}, {Mehl}, {Meyer}, {Millea}, {Mocanu}, {Mohr}, {Montroy}, {Padin},
  {Plagge}, {Pryke}, {Ruhl}, {Schaffer}, {Shaw}, {Shirokoff}, {Spieler},
  {Staniszewski}, {Stark}, {Story}, {van Engelen}, {Vanderlinde}, {Vieira},
  {Williamson}, \& {Zahn}}]{george15}
{George}, E.~M., {et~al.} 2015, \apj, 799, 177

\bibitem[{{Giannantonio} {et~al.}(2016){Giannantonio}, {Fosalba}, {Cawthon},
  {Omori}, {Crocce}, {Elsner}, {Leistedt}, {Dodelson}, {Benoit-L{\'e}vy},
  {Gazta{\~n}aga}, {Holder}, {Peiris}, {Percival}, {Kirk}, {Bauer}, {Benson},
  {Bernstein}, {Carretero}, {Crawford}, {Crittenden}, {Huterer}, {Jain},
  {Krause}, {Reichardt}, {Ross}, {Simard}, {Soergel}, {Stark}, {Story},
  {Vieira}, {Weller}, {Abbott}, {Abdalla}, {Allam}, {Armstrong}, {Banerji},
  {Bernstein}, {Bertin}, {Brooks}, {Buckley-Geer}, {Burke}, {Capozzi},
  {Carlstrom}, {Carnero Rosell}, {Carrasco Kind}, {Castander}, {Chang},
  {Cunha}, {da Costa}, {D'Andrea}, {DePoy}, {Desai}, {Diehl}, {Dietrich},
  {Doel}, {Eifler}, {Evrard}, {Neto}, {Fernandez}, {Finley}, {Flaugher},
  {Frieman}, {Gerdes}, {Gruen}, {Gruendl}, {Gutierrez}, {Holzapfel},
  {Honscheid}, {James}, {Kuehn}, {Kuropatkin}, {Lahav}, {Li}, {Lima}, {March},
  {Marshall}, {Martini}, {Melchior}, {Miquel}, {Mohr}, {Nichol}, {Nord},
  {Ogando}, {Plazas}, {Romer}, {Roodman}, {Rykoff}, {Sako}, {Saliwanchik},
  {Sanchez}, {Schubnell}, {Sevilla-Noarbe}, {Smith}, {Soares-Santos},
  {Sobreira}, {Suchyta}, {Swanson}, {Tarle}, {Thaler}, {Thomas}, {Vikram},
  {Walker}, {Wechsler}, \& {Zuntz}}]{giannantonio16}
{Giannantonio}, T., {et~al.} 2016, \mnras, 456, 3213

\bibitem[{{G{\'o}rski} {et~al.}(2005){G{\'o}rski}, {Hivon}, {Banday},
  {Wandelt}, {Hansen}, {Reinecke}, \& {Bartelmann}}]{gorski05}
{G{\'o}rski}, K.~M., {Hivon}, E., {Banday}, A.~J., {Wandelt}, B.~D., {Hansen},
  F.~K., {Reinecke}, M., \& {Bartelmann}, M. 2005, \apj, 622, 759

\bibitem[{{Hanson} {et~al.}(2011){Hanson}, {Challinor}, {Efstathiou}, \&
  {Bielewicz}}]{hanson11}
{Hanson}, D., {Challinor}, A., {Efstathiou}, G., \& {Bielewicz}, P. 2011, \prd,
  83, 043005

\bibitem[{{Hoffman} \& {Ribak}(1991)}]{hoffman91}
{Hoffman}, Y., \& {Ribak}, E. 1991, \apjl, 380, L5

\bibitem[{{Hou} {et~al.}(2017){Hou}, {Aylor}, {Benson}, {Bleem}, {Carlstrom},
  {Chang}, {Cho}, {Chown}, {Crawford}, {Crites}, {de Haan}, {Dobbs}, {Everett},
  {Follin}, {George}, {Halverson}, {Harrington}, {Holder}, {Holzapfel},
  {Hrubes}, {Keisler}, {Knox}, {Lee}, {Leitch}, {Luong-Van}, {Marrone},
  {McMahon}, {Meyer}, {Millea}, {Mocanu}, {Mohr}, {Natoli}, {Omori}, {Padin},
  {Pryke}, {Reichardt}, {Ruhl}, {Sayre}, {Schaffer}, {Shirokoff},
  {Staniszewski}, {Stark}, {Story}, {Vanderlinde}, {Vieira}, \&
  {Williamson}}]{hou17}
{Hou}, Z., {et~al.} 2017, ArXiv e-prints, 1704.00884

\bibitem[{{Howlett} {et~al.}(2012){Howlett}, {Lewis}, {Hall}, \&
  {Challinor}}]{howlett12}
{Howlett}, C., {Lewis}, A., {Hall}, A., \& {Challinor}, A. 2012, \jcap, 4, 027

\bibitem[{{Hu} \& {Okamoto}(2002)}]{hu02a}
{Hu}, W., \& {Okamoto}, T. 2002, \apj, 574, 566

\bibitem[{Hunter(2007)}]{hunter07}
Hunter, J.~D. 2007, Computing In Science \& Engineering, 9, 90

\bibitem[{{Huterer} {et~al.}(2015){Huterer}, {Kirkby}, {Bean}, {Connolly},
  {Dawson}, {Dodelson}, {Evrard}, {Jain}, {Jarvis}, {Linder}, {Mandelbaum},
  {May}, {Raccanelli}, {Reid}, {Rozo}, {Schmidt}, {Sehgal}, {Slosar}, {van
  Engelen}, {Wu}, \& {Zhao}}]{huterer15}
{Huterer}, D., {et~al.} 2015, Astroparticle Physics, 63, 23

\bibitem[{{Kaiser}(1992)}]{kaiser92}
{Kaiser}, N. 1992, \apj, 388, 272

\bibitem[{{Kesden} {et~al.}(2002){Kesden}, {Cooray}, \&
  {Kamionkowski}}]{kesden02}
{Kesden}, M., {Cooray}, A., \& {Kamionkowski}, M. 2002, Physical Review
  Letters, 89, 011304 (4 pages)

\bibitem[{{Kesden} {et~al.}(2003){Kesden}, {Cooray}, \&
  {Kamionkowski}}]{kesden03}
------. 2003, \prd, 67, 123507

\bibitem[{{Kilbinger}(2015)}]{kilbinger15}
{Kilbinger}, M. 2015, Reports on Progress in Physics, 78, 086901

\bibitem[{{Kirk} {et~al.}(2015){Kirk}, {Hilton}, {Cress}, {Crawford}, {Hughes},
  {Battaglia}, {Bond}, {Burke}, {Gralla}, {Hajian}, {Hasselfield}, {Hincks},
  {Infante}, {Kosowsky}, {Marriage}, {Menanteau}, {Moodley}, {Niemack},
  {Sievers}, {Sif{\'o}n}, {Wilson}, {Wollack}, \& {Zunckel}}]{kirk15}
{Kirk}, B., {et~al.} 2015, \mnras, 449, 4010

\bibitem[{{Kirk} {et~al.}(2016){Kirk}, {Omori}, {Benoit-L{\'e}vy}, {Cawthon},
  {Chang}, {Larsen}, {Amara}, {Bacon}, {Crawford}, {Dodelson}, {Fosalba},
  {Giannantonio}, {Holder}, {Jain}, {Kacprzak}, {Lahav}, {MacCrann}, {Nicola},
  {Refregier}, {Sheldon}, {Story}, {Troxel}, {Vieira}, {Vikram}, {Zuntz},
  {Abbott}, {Abdalla}, {Becker}, {Benson}, {Bernstein}, {Bernstein}, {Bleem},
  {Bonnett}, {Bridle}, {Brooks}, {Buckley-Geer}, {Burke}, {Capozzi},
  {Carlstrom}, {Rosell}, {Kind}, {Carretero}, {Crocce}, {Cunha}, {D'Andrea},
  {da Costa}, {Desai}, {Diehl}, {Dietrich}, {Doel}, {Eifler}, {Evrard},
  {Flaugher}, {Frieman}, {Gerdes}, {Goldstein}, {Gruen}, {Gruendl},
  {Honscheid}, {James}, {Jarvis}, {Kent}, {Kuehn}, {Kuropatkin}, {Lima},
  {March}, {Martini}, {Melchior}, {Miller}, {Miquel}, {Nichol}, {Ogando},
  {Plazas}, {Reichardt}, {Roodman}, {Rozo}, {Rykoff}, {Sako}, {Sanchez},
  {Scarpine}, {Schubnell}, {Sevilla-Noarbe}, {Simard}, {Smith},
  {Soares-Santos}, {Sobreira}, {Suchyta}, {Swanson}, {Tarle}, {Thomas},
  {Wechsler}, \& {Weller}}]{kirk16}
{Kirk}, D., {et~al.} 2016, \mnras, 459, 21

\bibitem[{{Lewis}(2005)}]{lewis05}
{Lewis}, A. 2005, \prd, 71, 083008

\bibitem[{{Lewis} \& {Challinor}(2006)}]{lewis06}
{Lewis}, A., \& {Challinor}, A. 2006, \physrep, 429, 1

\bibitem[{{Lewis} {et~al.}(2000){Lewis}, {Challinor}, \& {Lasenby}}]{lewis00}
{Lewis}, A., {Challinor}, A., \& {Lasenby}, A. 2000, \apj, 538, 473

\bibitem[{{Namikawa} {et~al.}(2013){Namikawa}, {Hanson}, \&
  {Takahashi}}]{namikawa13}
{Namikawa}, T., {Hanson}, D., \& {Takahashi}, R. 2013, \mnras, 431, 609

\bibitem[{{Namikawa} {et~al.}(2012){Namikawa}, {Yamauchi}, \&
  {Taruya}}]{namikawa12}
{Namikawa}, T., {Yamauchi}, D., \& {Taruya}, A. 2012, \jcap, 1, 007

\bibitem[{{Okamoto} \& {Hu}(2003)}]{okamoto03}
{Okamoto}, T., \& {Hu}, W. 2003, \prd, 67, 083002

\bibitem[{{Osborne} {et~al.}(2014){Osborne}, {Hanson}, \&
  {Dor{\'e}}}]{osborne14}
{Osborne}, S.~J., {Hanson}, D., \& {Dor{\'e}}, O. 2014, \jcap, 3, 024

\bibitem[{{Planck Collaboration} {et~al.}(2015{\natexlab{a}}){Planck
  Collaboration}, {Adam}, {Ade}, {Aghanim}, {Akrami}, {Alves}, {Arnaud},
  {Arroja}, {Aumont}, {Baccigalupi}, \& et~al.}]{planck15-1}
{Planck Collaboration}, {et~al.} 2015{\natexlab{a}}, ArXiv e-prints, 1502.01582

\bibitem[{{Planck Collaboration} {et~al.}(2016){Planck Collaboration}, {Adam},
  {Ade}, {Aghanim}, {Arnaud}, {Ashdown}, {Aumont}, {Baccigalupi}, {Banday},
  {Barreiro}, \& et~al.}]{planck15-8}
------. 2016, \aap, 594, A8

\bibitem[{{Planck Collaboration} {et~al.}(2014){Planck Collaboration}, {Ade},
  {Aghanim}, {Armitage-Caplan}, {Arnaud}, {Ashdown}, {Atrio-Barandela},
  {Aumont}, {Baccigalupi}, {Banday}, \& et~al.}]{planck13-18}
------. 2014, \aap, 571, A18

\bibitem[{{Planck Collaboration} {et~al.}(2015{\natexlab{b}}){Planck
  Collaboration}, {Ade}, {Aghanim}, {Arnaud}, {Ashdown}, {Aumont},
  {Baccigalupi}, {Banday}, {Barreiro}, {Bartlett}, \& et~al.}]{planck15-12}
------. 2015{\natexlab{b}}, ArXiv e-prints, 1509.06348

\bibitem[{{Planck Collaboration} {et~al.}(2015{\natexlab{c}}){Planck
  Collaboration}, {Ade}, {Aghanim}, {Arnaud}, {Ashdown}, {Aumont},
  {Baccigalupi}, {Banday}, {Barreiro}, {Bartlett}, \& et~al.}]{planck15-15}
------. 2015{\natexlab{c}}, ArXiv e-prints, 1502.01591

\bibitem[{{Planck Collaboration} {et~al.}(2015{\natexlab{d}}){Planck
  Collaboration}, {Aghanim}, {Arnaud}, {Ashdown}, {Aumont}, {Baccigalupi},
  {Banday}, {Barreiro}, {Bartlett}, {Bartolo}, \& et~al.}]{planck15-11}
------. 2015{\natexlab{d}}, ArXiv e-prints, 1507.02704

\bibitem[{{POLARBEAR Collaboration} {et~al.}(2013){POLARBEAR Collaboration},
  {Ade}, {Akiba}, {Anthony}, {Arnold}, {Atlas}, {Barron}, {Boettger},
  {Borrill}, {Chapman}, {Chinone}, {Dobbs}, {Elleflot}, {Errard}, {Fabbian},
  {Feng}, {Flanigan}, {Gilbert}, {Grainger}, {Halverson}, {Hasegawa},
  {Hattori}, {Hazumi}, {Holzapfel}, {Hori}, {Howard}, {Hyland}, {Inoue},
  {Jaehnig}, {Jaffe}, {Keating}, {Kermish}, {Keskitalo}, {Kisner}, {Le Jeune},
  {Lee}, {Linder}, {Leitch}, {Lungu}, {Matsuda}, {Matsumura}, {Meng}, {Miller},
  {Morii}, {Moyerman}, {Myers}, {Navaroli}, {Nishino}, {Paar}, {Peloton},
  {Quealy}, {Rebeiz}, {Reichardt}, {Richards}, {Ross}, {Schanning}, {Schenck},
  {Sherwin}, {Shimizu}, {Shimmin}, {Shimon}, {Siritanasak}, {Smecher},
  {Spieler}, {Stebor}, {Steinbach}, {Stompor}, {Suzuki}, {Takakura}, {Tomaru},
  {Wilson}, {Yadav}, \& {Zahn}}]{polarbear13b}
{POLARBEAR Collaboration}, {et~al.} 2013, ArXiv e-prints, 1312.6646

\bibitem[{Pérez \& Granger(2007)}]{ipython}
Pérez, F., \& Granger, B.~E. 2007, Computing in Science \& Engineering, 9, 21

\bibitem[{{Schaffer} {et~al.}(2011){Schaffer}, {Crawford}, {Aird}, {Benson},
  {Bleem}, {Carlstrom}, {Chang}, {Cho}, {Crites}, {de Haan}, {Dobbs}, {George},
  {Halverson}, {Holder}, {Holzapfel}, {Hoover}, {Hrubes}, {Joy}, {Keisler},
  {Knox}, {Lee}, {Leitch}, {Lueker}, {Luong-Van}, {McMahon}, {Mehl}, {Meyer},
  {Mohr}, {Montroy}, {Padin}, {Plagge}, {Pryke}, {Reichardt}, {Ruhl},
  {Shirokoff}, {Spieler}, {Stalder}, {Staniszewski}, {Stark}, {Story},
  {Vanderlinde}, {Vieira}, \& {Williamson}}]{schaffer11}
{Schaffer}, K.~K., {et~al.} 2011, \apj, 743, 90

\bibitem[{{Shaw} {et~al.}(2010){Shaw}, {Nagai}, {Bhattacharya}, \&
  {Lau}}]{shaw10}
{Shaw}, L.~D., {Nagai}, D., {Bhattacharya}, S., \& {Lau}, E.~T. 2010, \apj,
  725, 1452

\bibitem[{{Shaw} {et~al.}(2012){Shaw}, {Rudd}, \& {Nagai}}]{shaw12}
{Shaw}, L.~D., {Rudd}, D.~H., \& {Nagai}, D. 2012, \apj, 756, 15

\bibitem[{{Sherwin} {et~al.}(2016){Sherwin}, {van Engelen}, {Sehgal},
  {Madhavacheril}, {Addison}, {Aiola}, {Allison}, {Battaglia}, {Beall},
  {Becker}, {Bond}, {Calabrese}, {Datta}, {Devlin}, {Dunner}, {Dunkley}, {Fox},
  {Gallardo}, {Halpern}, {Hasselfield}, {Henderson}, {Hill}, {Hilton},
  {Hubmayr}, {Hughes}, {Hincks}, {Hlozek}, {Huffenberger}, {Koopman},
  {Kosowsky}, {Louis}, {Maurin}, {McMahon}, {Moodley}, {Naess}, {Nati},
  {Newburgh}, {Niemack}, {Page}, {Sievers}, {Spergel}, {Staggs}, {Thornton},
  {Van Lanen}, {Vavagiakis}, \& {Wollack}}]{sherwin16}
{Sherwin}, B.~D., {et~al.} 2016, ArXiv e-prints, 1611.09753

\bibitem[{{Story} {et~al.}(2015){Story}, {Hanson}, {Ade}, {Aird}, {Austermann},
  {Beall}, {Bender}, {Benson}, {Bleem}, {Carlstrom}, {Chang}, {Chiang}, {Cho},
  {Citron}, {Crawford}, {Crites}, {de Haan}, {Dobbs}, {Everett}, {Gallicchio},
  {Gao}, {George}, {Gilbert}, {Halverson}, {Harrington}, {Henning}, {Hilton},
  {Holder}, {Holzapfel}, {Hoover}, {Hou}, {Hrubes}, {Huang}, {Hubmayr},
  {Irwin}, {Keisler}, {Knox}, {Lee}, {Leitch}, {Li}, {Liang}, {Luong-Van},
  {McMahon}, {Mehl}, {Meyer}, {Mocanu}, {Montroy}, {Natoli}, {Nibarger},
  {Novosad}, {Padin}, {Pryke}, {Reichardt}, {Ruhl}, {Saliwanchik}, {Sayre},
  {Schaffer}, {Smecher}, {Stark}, {Tucker}, {Vanderlinde}, {Vieira}, {Wang},
  {Whitehorn}, {Yefremenko}, \& {Zahn}}]{story15}
{Story}, K.~T., {et~al.} 2015, \apj, 810, 50

\bibitem[{{Story} {et~al.}(2013){Story}, {Reichardt}, {Hou}, {Keisler}, {Aird},
  {Benson}, {Bleem}, {Carlstrom}, {Chang}, {Cho}, {Crawford}, {Crites}, {de
  Haan}, {Dobbs}, {Dudley}, {Follin}, {George}, {Halverson}, {Holder},
  {Holzapfel}, {Hoover}, {Hrubes}, {Joy}, {Knox}, {Lee}, {Leitch}, {Lueker},
  {Luong-Van}, {McMahon}, {Mehl}, {Meyer}, {Millea}, {Mohr}, {Montroy},
  {Padin}, {Plagge}, {Pryke}, {Ruhl}, {Sayre}, {Schaffer}, {Shaw}, {Shirokoff},
  {Spieler}, {Staniszewski}, {Stark}, {van Engelen}, {Vanderlinde}, {Vieira},
  {Williamson}, \& {Zahn}}]{story13}
------. 2013, \apj, 779, 86

\bibitem[{{Szapudi} {et~al.}(2001){Szapudi}, {Prunet}, {Pogosyan}, {Szalay}, \&
  {Bond}}]{szapudi01}
{Szapudi}, I., {Prunet}, S., {Pogosyan}, D., {Szalay}, A.~S., \& {Bond}, J.~R.
  2001, \apjl, 548, L115

\bibitem[{van~der Walt {et~al.}(2011)van~der Walt, Colbert, \&
  Varoquaux}]{numpy}
van~der Walt, S., Colbert, S.~C., \& Varoquaux, G. 2011, Computing in Science
  \& Engineering, 13, 22

\bibitem[{{van Engelen} {et~al.}(2014){van Engelen}, {Bhattacharya}, {Sehgal},
  {Holder}, {Zahn}, \& {Nagai}}]{vanengelen14}
{van Engelen}, A., {Bhattacharya}, S., {Sehgal}, N., {Holder}, G.~P., {Zahn},
  O., \& {Nagai}, D. 2014, \apj, 786, 13

\bibitem[{{van Engelen} {et~al.}(2012){van Engelen}, {Keisler}, {Zahn}, {Aird},
  {Benson}, {Bleem}, {Carlstrom}, {Chang}, {Cho}, {Crawford}, {Crites}, {de
  Haan}, {Dobbs}, {Dudley}, {George}, {Halverson}, {Holder}, {Holzapfel},
  {Hoover}, {Hou}, {Hrubes}, {Joy}, {Knox}, {Lee}, {Leitch}, {Lueker},
  {Luong-Van}, {McMahon}, {Mehl}, {Meyer}, {Millea}, {Mohr}, {Montroy},
  {Natoli}, {Padin}, {Plagge}, {Pryke}, {Reichardt}, {Ruhl}, {Sayre},
  {Schaffer}, {Shaw}, {Shirokoff}, {Spieler}, {Staniszewski}, {Stark}, {Story},
  {Vanderlinde}, {Vieira}, \& {Williamson}}]{vanengelen12}
{van Engelen}, A., {et~al.} 2012, \apj, 756, 142

\bibitem[{{Weinberg} {et~al.}(2013){Weinberg}, {Mortonson}, {Eisenstein},
  {Hirata}, {Riess}, \& {Rozo}}]{weinberg13}
{Weinberg}, D.~H., {Mortonson}, M.~J., {Eisenstein}, D.~J., {Hirata}, C.,
  {Riess}, A.~G., \& {Rozo}, E. 2013, \physrep, 530, 87

\bibitem[{{Wright} {et~al.}(2010){Wright}, {Eisenhardt}, {Mainzer}, {Ressler},
  {Cutri}, {Jarrett}, {Kirkpatrick}, {Padgett}, {McMillan}, {Skrutskie},
  {Stanford}, {Cohen}, {Walker}, {Mather}, {Leisawitz}, {Gautier}, {McLean},
  {Benford}, {Lonsdale}, {Blain}, {Mendez}, {Irace}, {Duval}, {Liu}, {Royer},
  {Heinrichsen}, {Howard}, {Shannon}, {Kendall}, {Walsh}, {Larsen}, {Cardon},
  {Schick}, {Schwalm}, {Abid}, {Fabinsky}, {Naes}, \& {Tsai}}]{wright10}
{Wright}, E.~L., {et~al.} 2010, \aj, 140, 1868

\bibitem[{{Zaldarriaga} \& {Seljak}(1999)}]{zaldarriaga99}
{Zaldarriaga}, M., \& {Seljak}, U. 1999, \prd, 59, 123507

\end{thebibliography}

\onecolumngrid
\section{Appendix}

\end{document}